\documentclass[a4paper,11pt]{article}
\pdfoutput=1 

\usepackage{jheppub} 

\usepackage[T1]{fontenc} 

\def\beq{\begin{equation}}
\def\eeq{\end{equation}}
\def\bea{\begin{eqnarray}}
\def\eea{\end{eqnarray}}
\def\ba{\begin{array}}
\def\ea{\end{array}}
\def\bec{\begin{center}}
\def\ec{\end{center}}

\newcommand{\dis}[1]{\begin{equation}\begin{split}#1\end{split}\end{equation}}

\newcommand{\mplanck}{{M_{\rm Pl}}}

\newcommand{\gev}{\,\textrm{GeV}}

\newcommand{\Omsusy}{{{\cal O}(m_{3/2})}}
\newcommand{\Omsusysq}{{{\cal O}(m_{3/2}^2)}}
\newcommand{\msusy}{{m_{3/2}}}
\newcommand{\vpq}{{v_{\rm PQ}}}

\title{\boldmath Peccei-Quinn invariant singlet extended SUSY \\ with anomalous  $U(1)$ gauge symmetry }

\author[]{Sang Hui Im}
\author[]{and Min-Seok Seo }

\affiliation[]{ Center for Theoretical Physics of the Universe, Institute for Basic Science (IBS),\\ Daejeon 305-811, Korea}

\emailAdd{shim@ibs.re.kr}
\emailAdd{minseokseo@ibs.re.kr}

\abstract{Recent discovery of the SM-like Higgs boson with $m_h\simeq 125$ GeV motivates an extension of the minimal supersymmetric standard model (MSSM), which involves  a singlet Higgs superfield with a sizable Yukawa coupling to the doublet Higgs superfields. 
We examine such singlet-extended SUSY models with a Peccei-Quinn (PQ) symmetry that originates from an anomalous $U(1)_A$ gauge symmetry. We focus on the specific scheme that the PQ symmetry is spontaneously broken 
at an intermediate scale $v_{\rm PQ}\sim \sqrt{m_{\rm SUSY}M_{\rm Pl}}$ by an interplay between Planck scale suppressed operators and
tachyonic soft scalar mass $m_{\rm SUSY}\sim \sqrt{D_A}$ induced dominantly by the $U(1)_A$  $D$-term  $D_A$. This scheme also results in  spontaneous SUSY breaking in the PQ sector, generating the gaugino masses $M_{1/2}\sim \sqrt{D_A}$ when it is transmitted to the MSSM sector by the conventional gauge mediation mechanism.    
As a result, the  MSSM soft parameters in this scheme are induced  mostly by the $U(1)_A$ $D$-term and the gauge mediated SUSY breaking from the PQ sector, so that
 the sparticle masses can be near the present experimental bounds without causing the SUSY flavor problem. 
 The scheme is severely constrained by the condition that a phenomenologically viable form of the low energy operators of the singlet and doublet Higgs superfields  is generated by the PQ breaking sector in a way similar to the Kim-Nilles solution of the $\mu$ problem, and the resulting Higgs mass parameters allow the electroweak symmetry breaking with small $\tan\beta$.
 We find two minimal models with  two singlet Higgs superfields, satisfying this condition with a relatively simple form of the PQ breaking sector, and briefly discuss some phenomenological aspects of the model. 
 
 \vskip 0.5cm
 
 CTPU-14-12}

\begin{document} 
\maketitle
\flushbottom

\section{Introduction}\label{Sec:Introduction}

Low energy supersymmetry (SUSY) \cite{SUSYreview1, SUSYreview2} and the QCD axion \cite{Axionreview1, Axionreview2} are compelling candidates for physics beyond the Standard Model (SM) as they not only solve the major fine-tuning problems of the SM, {\it i.e.} the gauge hierarchy problem and the strong CP problem, but also shed a light on different fundamental issues such as dark matter and unification. Furthermore there are several virtues of having both SUSY and axion together. For instance, the axion scale  can be determined by an interplay between SUSY breaking scalar mass $m_{\rm SUSY}$ and a Planck scale suppressed higher dimensional operator, which would generate an intermediate axion scale $v_{\rm PQ}\sim \sqrt{m_{\rm SUSY}M_{\rm Pl}}$ in a natural way \cite{Murayama:1992dj}.
The absence of a potentially too large bare $\mu$ term of the doublet Higgs superfields can be understood also by a Peccei-Quinn (PQ) symmetry, $U(1)_{\rm PQ}$ \cite{PecceiQuinn, Peccei:1977ur} for the QCD axion. Then a right size of the Higgs $\mu$ parameter can be generated by the spontaneous PQ breaking as $\mu\sim v_{\rm PQ}^2/M_{\rm Pl}\sim m_{\rm SUSY}$, solving the $\mu$ problem for the supersymmetric Higgs sector \cite{Kim:1983dt}. As another possible virtue, the cosmological PQ phase transition in such model can be preceded by  a thermal inflation, thereby  solves the cosmological moduli problem \cite{Lyth:1995ka, Choi:1996vz}.

In view of minimizing the fine-tuning for the electroweak symmetry breaking (EWSB), we are most interested in the case that sparticles, particularly the  stops, are as light as possible,
being close to the present experimental bounds \cite{Papucci:2011wy}.
On the other hand, to explain the recently discovered SM-like Higgs boson mass $m_h\simeq 125$ GeV within the framework of the minimal supersymmetric standard model (MSSM), the stops need to have a mass around multi-TeV or even heavier, which is well above the current direct search limit \cite{Hall:2011aa, Feng:2013tvd}.
A simple way to avoid this difficulty is to extend the MSSM by adding a singlet Higgs superfield $S$
which has the superpotential  coupling $\lambda SH_uH_d$ to the doublet Higgs superfields $H_{u,d}$ \cite{NMSSM, Ellwanger:2009dp, Barbieri:2006bg}.  In such singlet-extended  models, the SM-like Higgs boson mass receives a contribution $\delta m_h^2 = \lambda^2 m_Z^2 \sin^2 2\beta/(g_1^2+g_2^2)$ from the Higgs quartic coupling $|\lambda H_uH_d|^2$, and the stops can have a relatively light mass around (or below)  TeV, while being compatible with   $m_h\simeq 125$ GeV, if
$\lambda \sim 1$ and $\tan\beta = \langle H_u^0\rangle/\langle H_d^0\rangle \sim 1$. This is perhaps the most straightforward extension of the MSSM, minimizing the fine-tuning for the EWSB under the known experimental constraints.  

The model can be extended further  
by introducing a PQ symmetry \cite{Jeong:2011jk, Jeong:2012ma, Kim:2012az, Bae:2012am, Choi:2013lda}, to solve the strong CP problem,  together with a PQ sector which breaks the PQ symmetry  spontaneously at an intermediate scale $v_{\rm PQ}\sim 10^9-10^{12}$ GeV generated by $ \sqrt{m_{\rm SUSY}M_{\rm Pl}}$ without introducing
new bare mass parameters \cite{Murayama:1992dj, Gherghetta:1995jx, Bae:2014yta}.
One can arrange the model further, so that all the low energy mass parameters of the singlet-extended Higgs sector are  generated by the spontaneous PQ breaking, and have a value comparable to $m_{\rm SUSY}$
in a way similar to the Kim-Nilles mechanism \cite{Kim:1983dt} for the $\mu$ problem.

An important issue about the axion solution of the strong CP problem is the UV origin of the PQ symmetry which is required to be protected well from quantum gravity effects violating global symmetries in general \cite{Abbott:1989jw, Coleman:1989zu, Kallosh:1995hi, Banks:2010zn}. 
Note that to solve the strong CP problem, the explicit PQ breaking by quantum gravity effects should be negligible compared to the breaking by the QCD anomaly \cite{Barr:1992qq, Kamionkowski:1992mf, Holman:1992us}. 
For the UV origin of a PQ symmetry, an appealing possibility is that $U(1)_{\rm PQ}$ originates from an anomalous $U(1)_A$ gauge symmetry whose gauge boson gains a heavy mass near the Planck scale by the St\"uckelberg mechanism \cite{Barr:1985hk, Kim:1988dd, Svrcek:2006yi, Choi:2011xt, Honecker:2013mya, Choi:2014uaa}.
 Then, quantum gravity effects breaking $U(1)_{\rm PQ}$ can be exponentially suppressed.

In this paper we examine the SUSY breaking, as well as some of the 
phenomenological consequences, in singlet-extended SUSY models involving  a PQ symmetry which originates from an anomalous $U(1)_A$ gauge symmetry.
We are interested in the scheme to yield flavor conserving soft parameters 
which lead to the superparticle masses near the present experimental bounds, together with $m_h\simeq 125$ GeV which is largely due to the singlet superpotential term $\lambda SH_uH_d$ with $\lambda\sim 1$ and $\tan\beta\sim 1$.  
In the next section, we first discuss
  generic features of SUSY breaking in models with anomalous $U(1)_A$ gauge symmetry broken by the St\"uckelberg mechanism, while leaving a global PQ symmetry as a low energy remnant \cite{Choi:2014uaa}.
We then examine the specific scheme that the soft SUSY breaking parameters in the PQ breaking sector
are dominated by the $U(1)_A$ $D$-term contribution as  
\dis{
\epsilon \,\equiv \, \frac{m_{\rm MM}}{\sqrt{D_A}} \,\ll \,1,}
 where $m_{\rm MM}$ denotes the moduli (or equivalently gravity) mediated soft masses.
In this scheme, the PQ symmetry is  spontaneously broken at 
$v_{\rm PQ}\sim (\sqrt{D_A}M_{\rm Pl})^{1/2}$, or more generically  
$v_{\rm PQ}\sim (\sqrt{D_A}M_{\rm Pl}^n)^{1/(n+1)}$ ($n\geq 1$),
by an interplay between the $D$-term induced tachyonic scalar mass and a Planck scale suppressed operator.
A notable feature of this scheme is that it leads to a spontaneous SUSY breaking in the PQ breaking sector,
showing a hierarchical structure for vacuum expectation values by $\epsilon$.
This SUSY breaking in the PQ breaking sector can be transmitted to the MSSM sector by the conventional gauge mediation mechanism, yielding the gauge mediated soft masses:
\dis{
m_{\rm GM} \,\sim \,  \frac{g^2}{8\pi^2\epsilon}\sqrt{D_A}.}
We will focus on a scheme in which $\epsilon$ amounts to 
\dis{
\epsilon \,\sim\, \frac{g^2}{8\pi^2},}
for which the MSSM soft parameters are determined by the gauge mediated SUSY breaking from the PQ breaking sector and the $U(1)_A$ $D$-term, which are comparable to each other.

To complete the scheme, we need to generate a phenomenologically viable form of the low energy operators of  the
singlet and doublet Higgs superfields through the spontaneous PQ breaking as in the Kim-Nilles solution of the $\mu$ problem.
It turns out that the hierarchical pattern 
of the SUSY breaking $F$-components in the PQ breaking sector makes this non-trivial at least for a relatively simple form of the PQ breaking sector.
 In Sec. \ref{Sec:PQMSSMmodel}, we present two minimally viable models involving two singlet Higgs superfields and discuss some of the phenomenological consequences of the models.
One of the models is more interesting as it allows the limit that
the Higgs sector including the singlet Higgs is parametrically lighter than the other sector of the model, without causing further fine-tuning than the minimal fine-tuning for the EWSB.

\section{Features of PQ symmetry and soft terms with anomalous $U(1)_A$  }\label{Sec:Model}


\subsection{Peccei-Quinn symmetry and $D$-term mediation from an anomalous $U(1)_A$}\label{subsec:PQfromD}
 
   We begin with an observation that a large fraction of phenomenologically viable string compactifications involves an anomalous $U(1)_A$ gauge symmetry.
    An anomalous $U(1)_A$ gauge symmetry can be quantum mechanically consistent by the Green-Schwarz (GS) anomaly cancellation \cite{Green:1984sg}, which is implemented by introducing the axion-like field $a_p$, a zero mode of the higher-dimensional $p$-form gauge field.
 In the supersymmetric language, $a_p$ is a pseudoscalar component of a chiral multiplet for the GS modulus $T_A$.
  Then various supermultiplets transform under $U(1)_A$ as 
 \dis{U(1)_A:\quad V_A\to V_A+\Lambda+\Lambda^*,\quad 
 T_A\to T_A+\delta_{\rm GS}\Lambda, \quad \Phi_i \to e^{-2q_i \Lambda}\Phi_i,\label{Eq:U(1)Atrans}} 
 where $V_A$ is the $U(1)_A$ vector multiplet, $\Lambda$ is a chiral multiplet parametrizing $U(1)_A$ gauge transformation, and a coefficient $\delta_{\rm GS}$ is the $U(1)_A$-QCD-QCD anomaly coefficient  given by
\dis{
\label{Eq:GScoeff}
\delta_{\rm GS}=\frac{1}{8\pi^2}\sum_i q_i {\rm Tr}(T_c(\Phi_i)^2),}
for  $T_c(\Phi_i)$ denoting the color charge matrix of $\Phi_i$.
  From this, one finds that the K\"ahler potential
   \dis{K=K_0(t_A, t_b, t_k)+Z_i (t_A, t_b, t_k)\Phi_i^* e^{2q_iV_A}\Phi_i,\label{Eq:GSKaehler}} 
   depends on $T_A$ through a gauge invariant combination $t_A\equiv T_A+T_A^*-\delta_{\rm GS}V_A$.
 We note that the K\"ahler potential Eq. (\ref{Eq:GSKaehler}) in general contains other moduli $t_k \equiv T_k+T_k^*$, as well as a SUSY breaking modulus $t_b \equiv T_b+T_b^*$, which are not charged under $U(1)_A$.

 Let $2\eta^I=\{-\delta_{\rm GS}, 2q_i\Phi_i \}$ be the holomorphic Killing vector fields generating an infinitesimal $U(1)_A$ gauge transformation of chiral multiplets $\Phi_I=\{T_A, \Phi_i\}$.
 Then the gauge boson mass and $D$-term of the $U(1)_A$ multiplet of gauge coupling $g_A$ are given by 
 \dis{&M_A^2=2g_A^2\eta^I{\bar \eta}^{\bar J}\partial_I\partial_{\bar J}K= 2g_A^2(M_{\rm GS}^2+M_{\rm matter}^2),\\
& D_A=-\eta^I\partial_I K=\xi_{\rm FI}+\tilde{M}_{\rm matter}^2, }
respectively, where the GS modulus contribution
\dis{M_{\rm GS}^2=\frac{\delta_{\rm GS}^2}{4}\partial_{t_A}^2 K_0,\quad \xi_{\rm FI}=\frac{\delta_{\rm GS}}{2}\partial_{t_A} K_0,}
and the matter contribution
\dis{&M_{\rm matter}^2=\sum_i\Big(q_i^2Z_i-q_i\delta_{\rm GS}\partial_{t_A}Z_i+\Big(\frac{\delta_{\rm GS}}{2}\Big)^2\partial_{t_A}^2 Z_i\Big)\Phi_i^*e^{2q_iV_A}\Phi_i,
\\
&\tilde{M}_{\rm matter}^2=-\sum_i\Big(q_iZ_i-\frac{\delta_{\rm GS}}{2}\partial_{t_A}Z_i\Big)\Phi_i^*e^{2q_iV_A}\Phi_i,
}
are written separately. 

 If the underlying string compactification admits a supersymmetric solution with vanishing Fayet-Illiopoulos (FI) term $\xi_{\rm FI}$ \cite{Ibanez:2012zz, Cvetic:2001nr, Cremades:2002te, Honecker:2004kb, Villadoro:2006ia, Gmeiner:2008xq, Blumenhagen:2005ga, Blumenhagen:2005pm, Anderson:2009nt}\footnote{If it were not the case, we need $q_i|\Phi_i|^2  \sim  \xi_{\rm FI} \sim \delta_{\rm GS} M_{\rm pl}^2$ for vanishing $D_A$ in the supersymmetric limit. Then the Higgs mechanism contribution ($\sim \delta_{\rm GS} M_{\rm pl}^2$) dominates over the St\"uckelberg mechanism contribution ($\sim \delta_{\rm GS}^2 M_{\rm pl}^2$). This is not appropriate for our purpose to obtain global $U(1)_{\rm PQ}$ symmetry as a remnant of $U(1)_A$.}, matter fields $\Phi_i$ do not develop vacuum expectation values (VEVs) in the supersymmetric limit in order to make $D$-term vanish.
  When $\partial_{t_A}^2 K_0 \sim {\cal O}(1)$, the $U(1)_A$ gauge boson gains a mass $M_A \sim \delta_{\rm GS}M_{\rm Pl} \sim 10^{16}$ GeV by eating up an axion-like field $a_p\equiv \sqrt{2}{\rm Im}T_A$ in the GS modulus (St\"uckelberg mechanism), rather than a pseudoscalar in the matter (Higgs mechanism), {\it i.e.} $M_A^2\sim M_{\rm GS}^2 \gg M_{\rm matter}^2 \sim |\Phi_i|^2$.
After the massive vector field $\tilde A_\mu=(A_\mu,a_p)$ is integrated out, the low energy effective theory below $M_A$ involves a global PQ symmetry which can be identified  as the global part of $U(1)_A$ {\it without} the transformation of $a_p$ : 
\dis{
 U(1)_{\rm PQ}: \,\, \Phi_i \rightarrow e^{iq_i \beta}\Phi_i \quad (\beta={\rm constant}).
}
 Because $U(1)_{\rm PQ}$ differs from the global part of the genuine gauge symmetry $U(1)_A$ only by the absence of  the non-linear transformation of $a_p$, any quantum gravity effect which breaks  $U(1)_{\rm PQ}$ explicitly can be exponentially suppressed by $e^{-t_p}$, where $t_p$ is the volume modulus of the $p$-cycle which is dual to the zero mode $a_p$.
 The model then has a sensible limit that the PQ breaking quantum gravity effects are negligible enough for $U(1)_{\rm PQ}$ to solve the strong CP problem, although it requires an understanding of the dynamics to stabilize the volume modulus $t_p$ at a sufficiently large value \cite{Svrcek:2006yi}.

  Let us discuss the decoupling of the massive gauge boson in more detail.
 Since we expect $M_A \gg m_{3/2}$, the massive vector multiplet $V_A$ is integrated out in an almost supersymmetric way. 
 Then, $V_A$ is fixed by the superfield equation of motion 
 \dis{\frac{\partial K}{\partial V_A}\simeq  0,\label{Eq:FIcond}}
 in the supersymmetric limit.
 The scalar component of Eq. (\ref{Eq:FIcond}) provides a stabilization of $t_A$.
 On the other hand, if SUSY is mainly broken by the modulus $T_b$ such that $\partial_{t_b}^2K |F^{T_b}|^2\simeq  3|m_{3/2}|^2$, the D-component of Eq. (\ref{Eq:FIcond}) provides the $U(1)_A$ $D$-term VEV
 \dis{g_A^2D_A \simeq \frac{2}{\delta_{\rm GS}}\frac{\partial_{t_A}\partial_{t_b}^2K_0}{\partial_{t_A}^2K_0}|F^{T_b}|^2 \sim \Big(\frac{\partial_{t_A}\partial_{t_b}^2 K_0}{\partial_{t_A}^2 K_0 \partial_{t_b}^2 K_0}\Big)\frac{|m_{3/2}|^2}{\delta_{\rm GS}} \equiv \frac{\epsilon_1}{\delta_{\rm GS}}|m_{3/2}|^2, \label{Eq:Dtermval}}
 where $\epsilon_1$ parametrizes the sequestering between a SUSY breaking sector and an $U(1)_A$ sector, which means that $\epsilon_1=0$ in the fully sequestered case. 
 The same result is obtained by imposing the $U(1)_A$ invariance condition $\eta^I\partial_I(V_F+V_D)=0$ to vacuum values \cite{Choi:2006bh}.  
 Suppose the sequestering parameter $\epsilon_1$ is of order of $1/8\pi^2\sim \delta_{\rm GS}$, which is the case when $\epsilon_1$ represents a mixing between $t_b$ and $t_A$ through loop correction to $t_A$ in the K\"ahler potential, as  observed in APPENDIX \ref{Sec:appA}. 
Then we have $D_A\sim m_{3/2}^2$ and it constitutes soft scalar masses as $m_i^2=-q_i D_A$.

\subsection{Soft terms in the PQ and visible sectors}\label{subsec:softterms}
 
 As we have seen, at energy scale below $M_A$, we have the PQ symmetry as a remnant of $U(1)_A$. 
 In general, PQ-charged matters and the SM gauge fields are described by K\"ahler potential, superpotential, and gauge kinetic functions given by
   \dis{& K=K_0(t_b, t_A, t_k)+Z_i(t_b, t_A, t_k) \Phi_i^* e^{2q_i V_A}\Phi_i,
   \\
   & W=W_0(T_b, T_A, T_k)+\frac{1}{3!}\lambda_{ijk}(T_b, T_A, T_k)\Phi_i\Phi_j\Phi_k+\frac{1}{n!}\kappa_{i_1i_2..i_n}(T_b, T_A, T_k)\Phi_{i_1}\Phi_{i_2}\cdots\Phi_{i_n},
   \\
   & f_a=\gamma_a(T_b, T_k) +k_a T_A,}
 where subscript $a$ of $f_a$ runs over the SM gauge group components, $SU(3)_c,  SU(2)_L$ and $U(1)_Y$.
   Using basic supergravity(SUGRA) relations
     \dis{&m_{3/2}=e^{K/2}W,\quad F^I=-e^{K/2}K^{I{\bar J}}(D_J W)^*,\quad D_I W=W_I+K_I W,
 \\
 &V_F=K_{I{\bar J}}F^IF^{\bar J}-3e^K|W|^2,\quad V_D=\frac{g_A^2}{2}D_A^2,}
  we deduce soft terms
 \dis{{\cal L}_{\rm soft}=-\frac12 M_a\lambda_a\lambda_a-\frac12 m_i^2|\hat{\phi_i}|^2-\frac{1}{3!}A_{ijk}\hat{\lambda}_{ijk}\hat{\phi}_i\hat{\phi}_j\hat{\phi}_k-\frac{1}{n!}A_\kappa^{i_1i_2..i_n}\hat{\kappa}_{i_1i_2..i_n}\hat{\phi}_{i_1}\hat{\phi}_{i_2}\cdots\hat{\phi}_{i_n}, }
 given by
 \dis{&A_{ijk}=-F^I\partial_I\ln\Big(\frac{\lambda_{ijk}}{e^{-K_0}Z_iZ_jZ_k}\Big)+\frac12(\gamma_i+\gamma_j+\gamma_k)\frac{F^C}{C},
 \\
 &A_\kappa^{i_1i_2..i_n}=(n-3)\frac{F^C}{C}-F^I\partial_I\ln\Big(\frac{\kappa_{i_1i_2..i_n}}{e^{-nK_0/3}Z_{i_1}Z_{i_2}..Z_{i_n}}\Big),
 \\
 &m_i^2=\frac23V_F-F^IF^{\bar J}\partial_I\partial_{\bar J}\ln(e^{-K_0/3}Z_i)-(q_i+\eta^I\partial_I \ln Z_i)g_A^2 D_A - \frac14 \frac{\partial \gamma_i}{\partial \ln \mu}\frac{F^C}{C},
 \\
 &\frac{M_a}{g_a^2}=\frac12 F^I\partial_I f_a-\frac{1}{8\pi^2}\sum_i{\rm Tr}(T_a^2(\phi_i))F^I\partial_I\ln(e^{-K_0/3}Z_i)-\frac{b_a}{16\pi^2}\frac{F^C}{C},\label{Eq:soft}}
 where $\hat{\phi}_i$ denotes canonically normalized scalar component of the chiral multiplet $\Phi_i$ and $\hat{\lambda}, \hat{\kappa}$ are Yukawa couplings in this basis : 
  \dis{\hat{\lambda}_{ijk}=\frac{\lambda_{ijk}}{\sqrt{e^{-K_0}Z_iZ_jZ_k}},\quad\quad \hat{\kappa}_{i_1 i_2 .. i_n}=\frac{\kappa_{i_1i_2..i_n}}{\sqrt{e^{-nK_0/3}Z_{i_1}Z_{i_2}..Z_{i_n}}}.}
 Therefore, so far as gauge mediation is not concerned, we have three origins of soft terms :
 \begin{itemize}
 \item {\bf Moduli mediation (gravity mediation)} 
 
  When SUSY is mainly broken by the modulus $T_b$ satisfying $\partial_{t_b}^2K |F^{T_b}|^2\simeq  3|m_{3/2}|^2$, gravity mediation takes a form of moduli mediation \cite{Kaplunovsky:1993rd, Brignole:1993dj}. 
  Its effects on soft masses are parametrized by how much the SUSY breaking sector is sequestered from the visible sector :
  \dis{\epsilon_2 m_{3/2} \equiv F^{T_b}\partial_{t_b}\ln(e^{-K_0/3}Z_i)\quad{\rm and}\quad F^I\partial_I f_a.}
  We assume that these two are of the same order, $F^I\partial_I f_a \sim \epsilon_2 m_{3/2}$.\footnote{In fact, when $f_a = k_a T_A$ we need to consider $F^A$, which is estimated to be $F^A \simeq -e^{K/2}K^{T_A T_b^*} F_{T_b}^* \sim e^{K/2}(\partial_{t_A}\partial_{t_b}K_0/\partial_{t_A}^2 K_0)F^{T_b}$. Hence, $F^I\partial_I f_a$ is relevant to $\epsilon_1$, rather than $\epsilon_2$. However, since we will be focusing on the specific choice $\epsilon_1 \sim \epsilon_2 \sim 1/8\pi^2$ in the following discussion, our assumption here is acceptable.}
  
  \item {\bf Anomaly mediation}
  
   Anomaly mediation \cite{Randall:1998uk, Giudice:1998xp, Bagger:1999rd} is parametrized by the conformal compensator $C$, whose SUSY breaking effect is given by
 \dis{\frac{F^C}{C}=\frac13 K_IF^I+e^{K/2}W^*,}
 where we have taken the Einstein frame gauge $C=e^{K/6}$.
 
 \item {\bf $D$-term mediation}
 
  Soft scalar masses get contribution from $D$-term mediation \cite{Kawamura:1996wn, Binetruy:1996uv, Dvali:1996rj, Kawamura:1996bd, Dvali:1997sr, Murakami:2001hk, Higaki:2003ig, Kors:2004hz, Babu:2005ui, Dudas:2005vv, ArkaniHamed:1998nu, Barreiro:1998nd, GarciadelMoral:2005js, Villadoro:2005yq, Parameswaran:2007kf, Achucarro:2006zf,  Gallego:2008sv, Krippendorf:2009zza, Dudas:2007nz, Dudas:2008qf, Burgess:2003ic, Jockers:2005zy, Choi:2006bh, Choi:2011xt} , $ -q_i D_A \sim 8\pi^2\epsilon_1 m_{3/2}^2$.
  In the specific case of $\epsilon_1 \sim 1/8\pi^2$, we have $D_A \sim m_{3/2}^2$. 
      
 \end{itemize}
 
 In general, the moduli mediation (or gravity mediation) can cause the SUSY flavor problem without some non-trivial assumptions. Thus we will consider the situation that the moduli mediation is somewhat suppressed by some amount of sequestering $\epsilon_2$. 
 If the SUSY breaking modulus $T_b$ contacts with the PQ and visible sector through loop correction in the K\"ahler potential, it is plausible to take $\epsilon_2 \sim 1/8\pi^2$, as estimated in APPENDIX \ref{Sec:appA}. 
  In this case, soft scalar masses are dominated by $D$-term mediation of order of $m_{3/2}$, whereas $A$-terms and gaugino masses mainly come from moduli mediation :
   \dis{&A_{ijk}\sim F^{T_b}\partial_{t_b} \ln(e^{-K_0}{Z}_i{Z}_j{Z}_k)\sim \Big(\frac{\partial_{t_b}\ln(e^{-K_0/3}{ Z})}{\sqrt{\partial_{t_b}^2 K_0}}\Big)m_{3/2}\equiv \epsilon_2 m_{3/2},
\\
&A_\kappa^{i_1i_2..i_n}\sim F^{T_b}\partial_{t_b}\ln\Big(e^{-nK_0/3}Z_{i_1}Z_{i_2}..Z_{i_n}\Big)\sim \epsilon_2 m_{3/2}\quad\quad (n\ne 3),
\\
&\frac{M_a}{g_a^2}\sim \frac12 F^{T_b}\partial_{t_b}f_a-\frac{1}{8\pi^2}\sum_i{\rm Tr}(T_a^2(\phi_i))F^{T_b}\partial_{t_b}\ln(e^{-K_0/3}{Z}_i)\sim \frac{1}{8\pi^2} m_{3/2},}
as well as anomaly mediation.

   Concerning the anomaly mediation effects, first consider the case of $F^C/C \sim m_{3/2}$. 
   In this case, gaugino masses, coming from moduli and anomaly mediation, are of order of $(1/8\pi^2)m_{3/2}$.
   They are one-loop suppressed compared to $\sqrt{D_A}$ with $\epsilon_1 \sim 1/8\pi^2$, the main contribution to soft scalar masses.
   For gaugino masses of order of TeV, we have the spectrum for split SUSY \cite{Wells:2003tf, ArkaniHamed:2004fb, Giudice:2004tc, ArkaniHamed:2004yi} with soft scalar masses of order of 100 TeV \cite{Kors:2004hz, Babu:2005ui, Dudas:2005vv}.
   However, since we are interested in singlet-extended SUSY with a percent level fine-tuning, we look for the situation in which both soft scalar masses and gaugino masses are of the same order, around TeV scale.
   This is achieved in two ways : one is to take $m_{3/2}\sim 100$ TeV and two sequestering parameters satisfying $\epsilon_2^2 \sim 8\pi^2 \epsilon_1 \sim (1/8\pi^2)^2$.
   Another is to keep $\epsilon_1 \sim \epsilon_2 \sim 1/8\pi^2$ and introduce gauge mediation \cite{Dine:1981za, Dimopoulos:1981au, Dine:1981gu, Nappi:1982hm, AlvarezGaume:1981wy, Dimopoulos:1982gm, Dine:1994vc, Dine:1995ag, Giudice:1998bp} to give gaugino masses of order of $m_{3/2} \sim \sqrt{D_A} \sim$ TeV \cite{Dvali:1997sr}.
  The first way requires peculiar three loop order mixing $\epsilon_1 \sim (1/8\pi^2)^3$ between the SUSY breaking sector and $U(1)_A$ sector, so we will not pursue this possibility.
   On the other hand, as will be discussed in Sec. \ref{subsec:PQbreaking}, with our specific parameter choice $\epsilon_1 \sim \epsilon_2 \sim 1/8\pi^2$, we can realize the latter case by introducing a PQ-charged messenger, with the help of a one-loop suppressed $A$-term $A_\kappa^{i_1\cdots i_n}\sim \epsilon_2 m_{3/2}$ with $n\ne 3$. 
   In order to obtain such a one-loop suppressed $A$-term, we need to make anomaly mediation negligibly small, because anomaly mediation gives $A$-term with $n\ne 3$ of order of $F^C/C$.
    If $F^C/C$ is of order of $m_{3/2}$, $A$-term is of order of $m_{3/2}$ as well.
   Therefore, we need a model in which the SUSY breaking modulus $T_b$ has a no-scale structure \cite{Cremmer:1983bf, Ellis:1983ei, Ellis:1983sf, Ellis:1984bm, Lahanas:1986uc}, $K\simeq -3\ln t_b$ and superpotential is independent of $T_b$ at the leading order, to give $F^C/C \simeq (1/3)\partial_{t_b} K F^{T_b} +m_{3/2} \simeq 0$.
   For example, in the large volume scenario \cite{Balasubramanian:2005zx, Conlon:2005ki}, where the large volume modulus $T_b$ of Calabi-Yau (CY) 3-fold breaks SUSY mainly with a no-scale structure, $F^C/C$ is negligibly small. 
   Moreover, the coupling between $T_b$ and $T_A$ through the loop correction of the form of $(1/t_b^m)(t_A-\alpha_A \ln t_b)^2$ in the K\"ahler potential, where $m$ is some integer, gives $\epsilon_1 \sim \epsilon_2 \sim 1/8\pi^2$.
   Detailed calculation can be found in Refs. \cite{Choi:2010gm, Shin:2011uk}, and briefly described in APPENDIX \ref{Sec:appA}. 

  In summary, in the parameter space we are interested in, in which  $F^C/C$ is negligible and $\epsilon_1 \sim \epsilon_2 \sim 1/8\pi^2$, we have soft terms given by
  \dis{m_i^2 \simeq -q_i D_A + c_i m_{\rm GM}^2 \sim m_{3/2}^2, \quad A \sim \epsilon_2 m_{3/2} \sim \frac{1}{8\pi^2} m_{3/2}, \quad M_a \simeq  c_a m_{\rm GM} \sim m_{3/2},}
  where $ m_{\rm GM}$ is the gauge mediation contribution of the order of $m_{3/2}$, as will be discussed in the following subsection.
  Here, sfermion soft masses are dominantly given by D-term and gauge mediation, while $A$-terms mainly come from moduli mediation.
  D-term mediation contribution has two properties.
  First, by assigning flavor universal PQ charges, we can make sfermion soft masses flavor universal as well. 
  Second, PQ charge conservation implies that RG running effects on soft masses are negligible down to the messenger scale for gauge mediation.
  For this reason, sfermion soft masses are similar to those expected from general gauge mediation \cite{GGM, Meade:2008wd, Carpenter:2008wi}, in which each of slepton and squark soft masses is flavor universal, but relation between them depends on the PQ charge assignments and the type of gauge mediation.
  On the other hand, gaugino masses are dominantly given by gauge mediation.
  Detailed spectrum of gauginos depends on the type of gauge mediation as well, which is not relevant to following discussion.

\subsection{Soft-term-induced spontaneous Peccei-Quinn symmetry breaking and Gauge mediation}\label{subsec:PQbreaking}

 Since we are interested in the intermediate PQ breaking scale $v_{\rm PQ}$ obtained through an interplay between $m_{3/2}$ and some cutoff scale $M_*$ (generically either the Planck scale or the GUT scale), we consider a model similar to that discussed in Ref. \cite{Murayama:1992dj}. 
 In our setup, as a soft scalar mass squared $m_i^2$ from a $D$-term mediation is proportional to an PQ charge $q_i$, we can make some of scalar fields tachyonic by assigning a positive charge.
 In this regard, our setup provides a natural situation for PQ symmetry breaking through the scenario in Ref. \cite{Murayama:1992dj}.
 
 To begin with, let us consider a non-renormalizable superpotential for PQ charged chiral multiplets  $X$ and $Y$,
 \dis{W_{\rm PQ}=y\frac{X^{n+2}Y}{M_*^n}. \label{WPQ}}
 For this, we assign PQ charges to satisfy $(n+2)q_{X}+q_{Y}=0$. 
 Together with soft terms, a potential for scalars $X$ and $Y$ is given by
 \dis{V_{\rm PQ}(X, Y)&=\frac{|y|^2}{M_*^{2n}}|X|^{2(n+2)}+\frac{|y|^2}{M_*^{2n}}(n+2)^2|X|^{2(n+1)} |Y|^2
\\
&+m_{X}^2|X|^2+m_{Y}^2|Y|^2+\Big(yA\frac{X^{n+2}Y}{M_*^n}+{\rm h.c.}\Big).}
Since the sign of $q_X$ is opposite to that of $q_Y$, one can assign $q_{X}>0$ and $q_{Y}<0$ so that  $X$ is tachyonic whereas $Y$ is not, at the origin of field space. 
Then the PQ symmetry is broken as $X$ takes the VEV and it induces non-zero $Y$ VEV :
\dis{&\langle |X| \rangle \simeq \frac{1}{(n+2)^{1/2(n+1)}|y|^{1/(n+1)}} \sqrt[n+1]{ |m_{X}| M_*^n},
\\
&\langle Y \rangle \simeq -\frac{1}{(n+2)^{(n+2)/2(n+1)}y^{1/(n+1)}}\frac{A^*|m_{X}|}{(n+2)|m_{X}|^2+|m_{Y}|^2} \sqrt[n+1]{ |m_{X}| M_*^n} .\label{Eq:XYVEV}}
Note that the linear dependence of superpotential on $Y$ results in the $ Y $ VEV proportional to the $A$-term which is suppressed by $\epsilon_2$, so there appears a hierarchy between $X$ and $Y$ VEVs,
\dis{\Big|\frac{Y}{X}\Big|=\frac{|A||m_{X}|}{\sqrt{n+2}[(n+2)|m_{X}|^2+m_{Y}^2]}\sim\frac{|A|}{\sqrt{D_A}}\sim \epsilon_2.}
We identify the higher scale $X$ VEV  as the PQ scale $v_{\rm PQ}$.
As $|m_{X, Y}|\sim m_{3/2} \sim \sqrt{D_A}$, for $M_* =\mplanck$, we have
 \dis{v_{\rm PQ}\equiv \langle X \rangle \sim |y|^{-1/(n+1)} \sqrt[n+1]{ m_{3/2} M_*^n}= \left\{ \begin{array}{ll}
|y|^{-1/2}10^{10}\gev& n=1\\
|y|^{-1/3}10^{12-13}\gev&n=2 \\
|y|^{-1/4}10^{14}\gev & n=3
\end{array}\right. .}
On the other hand, for $M_*=M_{\rm GUT}$, 
 \dis{v_{\rm PQ} \sim \left\{ \begin{array}{ll}
|y|^{-1/2}10^{9}\gev& n=1\\
|y|^{-1/3}10^{11-12}\gev&n=2 \\
|y|^{-1/4}10^{13}\gev & n=3
\end{array}\right. .}
Therefore, regarding a bound $10^9\gev<v_{\rm PQ}<10^{12}\gev$,  we favor $n=1$ for $M_*=\mplanck$ and $n=1, 2$ for $M_*=M_{\rm GUT}$.

 The soft SUSY breaking terms also induce spontaneous SUSY breaking for the $X, Y$ sector with non-zero 
VEVs of $F^{X, Y}$ :
\dis{&\Big|\frac{F^{X}}{X}\Big|=\frac{|A||m_{X}|^2}{(n+2)|m_{X}|^2+m_{Y}^2}\sim A\sim \epsilon_2 m_{3/2},
\\
&\Big|\frac{F^{Y}}{Y}\Big|=\frac{(n+2)|m_{X}|^2+m_{Y}^2}{|A|}\sim\frac{D_A}{A}\sim \frac{1}{\epsilon_2}m_{3/2}.\label{Eq:PQhier}}
In short, we have $X \sim v_{\rm PQ}(1+ \epsilon_2m_{3/2}\theta^2)$ while
$Y \sim v_{\rm PQ}(\epsilon_2+ m_{3/2}\theta^2)$. 
 Here we emphasize that $F^{Y}/Y$ is enhanced by one loop factor compared to $m_{3/2}$ due to the small $Y$ VEV. 

Now we can use non-zero $F$-terms of $X$ and $Y$ to generate gauge mediation \cite{Nakayama:2012zc, Bae:2014efa} adopting the KSVZ axion model \cite{Kim:1979if, Shifman:1980}, in which a PQ breaking field couples to a vector-like quark pair. 
Since $F^{Y}/Y$ is enhanced compared to $F^{X}/X$, let us consider the case where $Y$ couples to a pair of fields which is vector-like under the SM gauge group,
\dis{W=Y\Psi \Psi^c.}
The $\Psi \Psi^c$ pair plays the role of messenger of the gauge mediation whose size is given by
\dis{m_{\rm GM}\equiv \frac{1}{8\pi^2}\Big|\frac{F^{Y}}{Y}\Big| \sim \frac{1}{8\pi^2}\frac{D_A}{A} \sim
 \frac{1}{8\pi^2\epsilon_2}\sqrt{D_A}. \label{Eq:mSUSY}}
 The gauge mediation gives gaugino masses comparable to soft scalar masses $\sim \sqrt{D_A}$
 for $\epsilon_2 \sim 1/8\pi^2$.
  We emphasize that a hierarchy between two SUSY breaking scales $A$ and  $\sqrt{D_A}$ and the accordingly induced $X$ and $Y$ VEV hierarchy are important aspects for our scheme to realize a low fine-tuned SUSY scenario.
 Especially the $X$ and $Y$ VEV hierarchy is obtained by the linear dependence of the superpotential on $Y$. In this regard, our choice of the superpotential (\ref{WPQ}) is quite generic.
 
\section{PQ invariant singlet extended SUSY models}\label{Sec:PQMSSMmodel}

In the previous section, we have specified the UV origin of a global $U(1)_{\rm PQ}$ symmetry and 
corresponding SUSY breaking mediation scheme which can realize all superparticle masses around TeV scale 
in order to minimize fine-tuning for EWSB without causing the SUSY flavor problem.
Now we will apply this scheme to singlet extended SUSY models at TeV scale to complete a low fine-tuned
SUSY scenario. 

We will consider singlet extended Higgs sector like the NMSSM models \cite{NMSSM, Ellwanger:2009dp} with $\Delta W = \lambda S H_u H_d $ coupling
to obtain the Higgs mass of 125 GeV through TeV scale SUSY, however, with possibly more than one singlet superfield in general.
Therefore, the general effective superpotential of the Higgs sector at TeV scale is given by\footnote{The bare $\mu$ and $B\mu$ parameters can be always shifted away by choosing an appropriate field basis
for $S_i$ at some scale, although there can be non-zero RG running contribution to $B\mu$ below the chosen scale.}     
\dis{W_\textrm{eff}=  \sum_{i} \lambda_i(1+\theta^2 A_i) S_i H_uH_d + f(S_i).}
For convenience, we define a singlet field $S_H$ by $\lambda S_H \equiv \sum_{i} \lambda_i S_i$. 
Then, in the field basis that the singlet fields are given by $S_H$ and its orthogonal fields, the general superpotential becomes
\dis{W_\textrm{eff}=\lambda(1+\theta^2 A_\lambda) S_H H_uH_d +\theta^2 \sum_{j} \lambda'_j A'_j S_j H_u H_d+f(S_H, S_j), \label{general W}}
where $S_j$ denotes the singlet fields orthogonal to $S_H$, which do not have a supersymmetric coupling to the doublet Higgs fields $H_u H_d$.
In this generalized Higgs sector, the effective $\mu$ and $B\mu$ parameters are given by
\dis{\mu_{\rm eff}=\lambda \langle S_H\rangle,\quad\quad 
(B\mu)_{\rm eff}=\lambda \langle \partial_{S_H} f^*\rangle +\lambda A_\lambda \langle S_H\rangle  + \sum_{j} \lambda'_j A'_j \langle S_j \rangle.}
The EWSB conditions in terms of these parameters are expressed as
\dis{ \label{ewcon_orig}
&\frac12 m_Z^2=\frac{m_{H_d}^2-m_{H_u}^2\tan^2\beta}{\tan^2\beta-1}-\mu_{\rm eff}^2,
\\
&\sin2\beta=\frac{2(B\mu)_{\rm eff}}{2\mu_{\rm eff}^2+m_{H_u}^2+m_{H_d}^2+\lambda^2v^2}.
}

From (\ref{ewcon_orig}), we find that the following condition must be fulfilled to realize a low fine-tuned EWSB :
\dis{ \label{ewcon_e}
100~\textrm{GeV} \lesssim \mu_\textrm{eff} \lesssim m_{H_{u, d}}  \lesssim 1~\textrm{TeV},~~ 
(B\mu)_\textrm{eff}  \sim m_{H_{u, d}}^2,}
where $ 1 $ TeV upper bound is set to achieve a small fine-tuning within a percent level \cite{Papucci:2011wy}. 
The lower bound for $ \mu_\textrm{eff} \gtrsim 100 $ GeV is imposed due to the LEP exclusion on chargino masses \cite{LEPchargino}. 
The remaining part of the condition is obtained as we require $\tan \beta$ close to 1 ($ \sin 2\beta \sim 1 $) to get a large tree level Higgs mass
beyond the MSSM,
because the Higgs quartic potential $\lambda^2 |H_uH_d|^2 $ from $ \Delta W=\lambda S_H H_u H_d $ is proportional to $ \sin ^2 2\beta $.


As discussed in Sec. \ref{Sec:Model}, soft scalar masses including $m_{H_{u, d}}$ are determined by $ \sqrt{D_A} $ and $m_{\rm GM} $ which are comparable to $m_{3/2}$.
Therefore, to satisfy the above condition (\ref{ewcon_e}), it is required that
$ \mu_\textrm{eff} \lesssim \Omsusy$ and
 $(B\mu)_\textrm{eff} \sim \Omsusysq $ with $m_{3/2} \sim $ 1 TeV.
Since the magnitude of $ \mu_\textrm{eff} $ and $ (B\mu)_\textrm{eff}  $ will be affected by the dimensionful singlet parameters in the singlet superpotential 
$f(S_H, S_j)$, the requirement means that those singlet parameters should be of $\Omsusy$. 
This can be achieved by PQ invariant higher dimensional operators.
For instance, we find a model that fulfills the condition (\ref{ewcon_e}) as follows:
\dis{ \label{eq:model1}
W_{\rm eff}=  \lambda S_{1} H_u H_d + \frac12 \kappa_1 S_1^2 S_2 + \frac12 \kappa_2 \frac{X^{n+1}}{M_*^n} S_2^2,
}
where $n =1$, 2, or 3 depending on the magnitude of PQ breaking scale $v_{\rm PQ}$. 
\begin{table}
\begin{center}
\begin{tabular}{| c || c | c | c | c |} 
\hline
    & $~~X~~$ &  $~~Y~~$  & $~~S_1~~$ & $~~S_2~~$ \\ \hline \hline
$ U(1)_{\textrm{PQ}}$ & $1$ & $-(n+2)$ & $(n+1)/4$ & $-(n+1)/2$  \\ \hline 
\end{tabular}  
\caption{Charge assignment for a minimal model \label{Table:model1}} 
\end{center}
\end{table}   
This model is given by the PQ charge assignment of Table \ref{Table:model1}.  
In Sec. \ref{subsec:PQbreaking}, we found that the PQ breaking fields $X, Y$ have VEVs as $ X \sim \vpq(1+\epsilon m_{3/2} \theta^2), Y \sim \vpq(\epsilon + m_{3/2} \theta^2)$. Therefore, by the relation $\vpq^{n+1}/M_*^n \sim m_{3/2}$, we obtain the singlet mass parameter
of ${\cal O}(m_{3/2})$ for the $S_2^2$ term. Then all dimensionful parameters of the singlet sector involving the soft masses
$m_{S_1}^2$ and $m_{S_2}^2$ are given around $m_{3/2}$, so the VEVs of $S_1$ and $S_2$ will be also of ${\cal O}(m_{3/2})$ if dimensionless couplings are of order unity. This gives $ \mu_\textrm{eff} $ and $ (B\mu)_\textrm{eff}  $ around $m_{3/2}$. 
A detailed analysis on the vacuum structure of this model will be discussed shortly. 

Notice that the model (\ref{eq:model1}) includes two singlet fields $S_1, S_2$. Actually it can be shown that at least two singlet fields are needed
to satisfy the condition (\ref{ewcon_e}) with a relatively simple PQ breaking sector.    
Moreover, we find that there are \emph{only two} working models with two singlet fields, one of which is (\ref{eq:model1}). 
The other model turns out to be
\dis{ \label{eq:model2}
W_{\rm eff} =  \lambda S_{2} H_u H_d + \frac12 \kappa_1 S_1^2 S_2 + \frac12 \kappa_2 \frac{X^{n+1}}{M_*^n} S_2^2.
}
This model has the same singlet sector as the model (\ref{eq:model1}).
The only difference between them is which singlet field ($S_1$ or $S_2$) couples to $H_u H_d$.

The main reasons why working models with two singlet fields are only those are related to the sequestering factor $\epsilon$ and the sign of PQ charges of singlet fields which determines the sign of 
their soft masses due to the $D$-term mediation.
Soft scalar masses $m_{S_1}^2=-q_{S_1}D_A$ and $m_{S_2}^2=-q_{S_2}D_A$ must have appropriate
sign depending on models for the singlet scalar fields to have non-zero VEVs,
and  the sequestering factor $\epsilon$ can make some necessary singlet parameters for a model too small below $\msusy$.
These severely constrain viable forms of models.
 The details to end up with the minimally viable models (\ref{eq:model1}) and (\ref{eq:model2}) 
can be found in APPENDIX \ref{Sec:appB}.  
 
Now let us discuss the vacuum structure and some phenomenological implications of the minimal models. 
Aside from $\lambda S_i H_u H_d$, the same singlet superpotential of the two models give a singlet scalar potential as
\dis{ \label{singlet_scalar_potential}
V_{\rm eff}(S_1, S_2) = & m_{S_1}^2|S_1|^2+(m_{S_2}^2+\mu_2'^2) |S_2|^2+\kappa_1^2|S_1|^2|S_2|^2
 \\
 &+\frac14\kappa_1^2|S_1|^4+\frac12\kappa_1 \mu_2' ({S_1^*}^2S_2+{\rm h.c.}),}
where $\mu_2' \equiv \kappa_2 \langle X \rangle^{n+1}/M_*^n$.
The soft masses $ m_{S_1}^2$ and $m_{S_2}^2$ are mainly induced from the $D$-term mediation while RG 
running effect is subdominant, so $ m_{S_1}^2<0$ and $m_{S_2}^2>0 $ from the PQ charges in Table \ref{Table:model1}. Hence $ S_1 $ gets  non-zero VEV with its quartic potential, while $ S_2 $ does so from its tadpole term after $ \langle S_1 \rangle $ becomes non-zero. Assuming $\mu_2'^2 (\sim \kappa_2^2 D_A) \lesssim m_{S_{1, 2}}^2 (\sim D_A)$ with $ \kappa_2 \lesssim {\cal O}(1) $, $S_1$ and $S_2$ get their VEVs as
 \dis{ \label{Eq:signletvevs}
 & |\langle S_1 \rangle|  \simeq\sqrt{\frac{-2m_{S_1}^2}{\kappa_1^2}}\sim \frac{|m_{S_1}|}{\kappa_1},
 \\
 & |\langle S_2 \rangle| =\frac12 \frac{\kappa_1 \mu_2'  |S_1| ^2}{m_{S_2}^2+\mu_2'^2+\kappa_1^2 |S_1|^2}\sim \frac{\mu_2'}{\kappa_1}.}
 
The parametric relations between the doublet Higgs sector and singlet sector are very different
depending on which singlet field couples to $H_u H_d$. For the model (\ref{eq:model1}) in which $ S_H = S_1 $, 
$\mu_\textrm{eff} =\lambda \langle S_H \rangle $ and $ (B\mu)_\textrm{eff} \simeq \lambda \langle \partial_{S_H} f \rangle$ are given by\footnote{The $A$-term contributions to $(B\mu)_\textrm{eff}$ can be neglected since they are suppressed by $\epsilon$ as discussed in Sec. \ref{Sec:Model}.} 
  \dis{ \label{model1mu}
  &\mu_{\rm eff} = \lambda\langle S_1 \rangle \sim \frac{m_{S_1}}{\kappa_1} \sim \frac{\sqrt{D_A}}{\kappa_1},
 \\
 &(B\mu)_{\rm eff}\simeq\lambda \kappa_1 \langle S_1S_2  \rangle \sim \frac{m_{S_1}\mu_2'}{\kappa_1} \sim \frac{\kappa_2}{\kappa_1}D_A.}
On the other hand, for the model (\ref{eq:model2}) in which $ S_H = S_2 $, we have
    \dis{ \label{model2mu}
    &\mu_{\rm eff} = \lambda \langle S_2 \rangle \sim \frac{\mu_2'}{\kappa_1}\sim \frac{\kappa_2}{\kappa_1}\sqrt{D_A},
    \\
    & (B\mu)_{\rm eff} \simeq \lambda\Big\langle\frac12\kappa_1 S_1^2 +\mu_2' S_2 \Big\rangle\sim \frac{m_{S_1}^2}{\kappa_1}\sim \frac{D_A}{\kappa_1}.}    
From the above relations, one can find that the dimensionless coefficients $ \kappa_1, \kappa_2 $ of the model (\ref{eq:model1}) should
be of $ {\cal O}(1) $ with $ \sqrt{D_A} \sim \msusy  $ to satisfy the conditions $ \mu_\textrm{eff} \lesssim \Omsusy$ and
 $(B\mu)_\textrm{eff} \sim \Omsusysq $. On the contrary, for the model (\ref{eq:model2}), we 
 observe that  $ \sqrt{D_A} $ smaller than $ \msusy$ is allowed even satisfying $ \mu_\textrm{eff} \lesssim \Omsusy$ and
 $(B\mu)_\textrm{eff} \sim \Omsusysq $ if $ \kappa_1, \kappa_2 $ are smaller than order unity.\footnote{ Smaller $ \sqrt{D_A}$ than $m_{3/2}$ can be obtained by assuming larger sequestering $\epsilon_1 \sim g^2/8\pi^2 < 1/8\pi^2$
between the $U(1)_A $ sector and SUSY breaking modulus in Eq. (\ref{Eq:Dtermval}). In this case, 
if $\epsilon_2$ is also of similar order with $\epsilon_1$, which is shown to be quite plausible in APPENDIX \ref{Sec:appA}, $m_\textrm{GM} $ in Eq. (\ref{Eq:mSUSY}) is still around $m_{3/2}$.} 
Since the singlet masses are governed by the scale of $ \sqrt{D_A} $, this means that
the singlet sector of the model (\ref{eq:model1}) must be around $ m_{3/2} \sim 1$ TeV similarly with the other SUSY sectors, while the singlet sector of the model (\ref{eq:model2}) can be parametrically lighter. 

A caveat must be placed, however, for the possibility of the relatively light singlet sector of the model (\ref{eq:model2}). Smaller $ \sqrt{D_A} $ than $ m_\textrm{GM} \sim m_{3/2} $ means that SUSY breaking is dominantly mediated by the gauge mediation.
If we take the {\it minimal} gauge mediation for the simplest case, it is known that $ \mu_\textrm{eff} $ cannot
be smaller than the scale of $ m_{H_u}(m_{\tilde t}) \sim m_{\tilde t} \sim m_{3/2} $ for the EWSB to occur (see Ref. \cite{Agashe:1999ct}, for instance). This can be problematic because of the mixing between the singlet
scalar $ S_H $ and the SM-like Higgs boson $ h $ :
\dis{ \label{mixing-hS_H}
m_{hS_H}^2 = \lambda v \left(2\mu_\textrm{eff} - (A_\lambda + \partial_{S_H}^2 f(S_1, S_2)) \sin 2\beta \right),  
} 
where $ f(S_1, S_2) $ is the singlet sector superpotential. For either $ S_H = S_1 $ or $ S_2$ , we find that $ \partial_{S_H}^2 f(S_1, S_2) \sim \mu_2' $ from (\ref{Eq:signletvevs}). 
Thus it will be around $ \lambda v \mu_\textrm{eff} $ unless there occurs some fine cancellation between $ \mu_\textrm{eff} $ and $ \mu_2' $.
To ensure the stability of the electroweak vacuum, the diagonal elements of the mass matrix must satisfy $ m_{hh}^2 m_{S_H S_H}^2 > m_{h S_H}^4 $ so that
 \dis{ \label{m_SHSH}
m_{S_H S_H}^2 \sim D_A \gtrsim \mu_\textrm{eff}^2.}
Therefore, relatively small $ \sqrt{D_A} $ requires also small $ \mu_\textrm{eff} $, which is 
impossible in the minimal gauge mediation. It means that, in the simplest case, the singlet sector is expected to be as heavy as the other SUSY sectors around $ \msusy \sim 1 $ TeV
for both models. 
Still, the gauge mediation can be realized more generally as in Refs. \cite{GGM, Meade:2008wd, Carpenter:2008wi} with small $ \mu_\textrm{eff} < \msusy$. A non-minimal gauge mediation, however, 
needs another SUSY breaking term comparable to $F_Y/Y \sim 16\pi^2 \msusy$ in Sec. \ref{subsec:PQbreaking}. It can arise from another copy of the spontaneous PQ breaking sector with $ X', Y'$, for example. Thus, the singlet sector in the model (\ref{eq:model2}) can be light but implies some complication of the model.

Let us more specifically describe the mass spectrum of the light singlet sector of 
the model (\ref{eq:model2}). From (\ref{model2mu}), we find that
$ \kappa_2 \lesssim \kappa_1 $ to satisfy $ \mu_\textrm{eff} \lesssim \Omsusy$ and
 $(B\mu)_\textrm{eff} \sim \Omsusysq $ as well as the condition (\ref{m_SHSH}). This gives
 \dis{
 \sqrt{D_A} &\sim \sqrt{\kappa_1} \msusy,~~ \mu_2' \sim \kappa_2\sqrt{\kappa_1}\msusy \lesssim \kappa_1^{3/2}
 \msusy, \\
 \mu_\textrm{eff} &\sim \frac{\kappa_2}{\sqrt{\kappa_1}}\msusy \lesssim \sqrt{\kappa_1}\msusy.
 }
Hence, for the coupling constants $ \kappa_2 \lesssim \kappa_1 < {\cal O}(1) $, there appear
hierarchical mass scales $ \mu_2' < \mu_\textrm{eff} \lesssim \sqrt{D_A} < \msusy $. 
For instance, $ \kappa_1 \sim 0.01 $ gives $ \sqrt{D_A} \sim 0.1 \msusy \sim {\cal O}(100) $ GeV and $ \mu_2' \lesssim {\cal O}(1) $ GeV.  Notice that the limit $ \mu_2' \rightarrow 0 $  corresponds to the PQ symmetric limit in which one pseudoscalar becomes massless,
and we find that the pseudoscalar mass is $ \sim \mu_2'^2 $ from the last term in the scalar potential (\ref{singlet_scalar_potential}).
Therefore, through the small coupling constants, we obtain a relatively light singlet scalar sector of their masses given by $ \sqrt{D_A} \sim \sqrt{\kappa_1} \msusy$ with an even lighter singlet pseudoscalar of mass
$ \mu_2' \lesssim \kappa_1^{3/2} \msusy $.

The singlet scalars have mixing with the doublet Higgs bosons. 
Assuming no cancellation between 
$ \mu_\textrm{eff} $ and $ \partial_{S_H}^2 f(S_1, S_2) \sim \mu_2' $ in (\ref{mixing-hS_H}), the mixing angle between the SM-like Higgs boson $ h $ and $ S_H $ is
\dis{
\theta_{hS_H} \simeq \frac{m^2_{hS_H}}{m^2_{S_HS_H} - m^2_{hh}} \sim
{\cal O}\left(\frac{\kappa_2}{\kappa_1^{3/2}}\frac{\lambda v}{m_{3/2}} \right) \lesssim {\cal O}\left(\frac{0.1}{\sqrt{\kappa_1}}\right),
} 
where $ \kappa_2 \lesssim \kappa_1 $ for the model (\ref{eq:model2}), while both $ \kappa_1 $ and $\kappa_2 $ should be of $ {\cal O}(1) $ for the model (\ref{eq:model1}). 
Similarly, the mixing angle between the SM-like Higgs boson $h$ and another scalar $S_j$ other than $ S_H $ is estimated to be
\dis{
\theta_{hS_j} \simeq \frac{m^2_{hS_j}}{m^2_{S_jS_j} - m^2_{hh}} &=
\frac{-\lambda v\partial_{S_j} \partial_{S_H} f(S_1, S_2) \sin 2\beta}{m^2_{S_jS_j} - m^2_{hh}}\\
&\sim
{\cal O}\left(\frac{1}{\sqrt{\kappa_1}}\frac{\lambda v}{m_{3/2}} \right) \sim {\cal O}\left(\frac{0.1}{\sqrt{\kappa_1}}\right),
} 
where $ \partial_{S_j} \partial_{S_H} f(S_1, S_2) = \partial_{S_1} \partial_{S_2} f(S_1, S_2) = \kappa_1 S_1 \sim m_{S_1} $ for either $ S_H = S_1 $ or $ S_2 $.
Therefore, the mixing angles can be quite sizable for the model (\ref{eq:model2}) with small $ \kappa_1 $
which results in departure from the SM Higgs boson properties with detectable signatures, while they are always as small as $ {\cal O}(0.1) $  for the model (\ref{eq:model1}).  
  
We briefly describe the neutralino sector. The singlino mass matrix in the basis of $(\tilde{S_1}, \tilde{S_2})$ is given by
\dis{M_{\rm singlino}=\left( \begin{array}{cc}
\kappa_1 \langle S_2  \rangle &\kappa_1 \langle S_1 \rangle\\
\kappa_1 \langle S_1 \rangle & \mu_2'
\end{array}\right) \sim \left( \begin{array}{cc}
\mu_2'  & m_{S_1}\\
m_{S_1} & \mu_2'
\end{array}\right),} 
and they mix with the doublet Higgsinos through the superpotential $ \lambda S_{H} H_u H_d $ with the
off-diagonal elements of $ {\cal O}(\lambda v) $. Thus the mixing angles between the doublet Higgsinos and singlinos are around ${\cal O}(\lambda v/ \sqrt{\kappa_1}\msusy) \lesssim  {\cal O}(0.1/\sqrt{\kappa_1})$.
When we assume the {\it minimal} gauge mediation, 
we have argued that all mass parameters $ m_{S_1}, \mu_2' $ and $ \mu_\textrm{eff} $
must be around $ \msusy \sim 1$ TeV with $ \kappa_1$, $ \kappa_2 $ of $ {\cal O}(1)$. In this case,
the Higgsinos and singlinos do not mix with each other so much, and they are all heavy around $ \msusy $ unless there occurs
fine cancellation between $ \mu_2' $ and $ m_{S_1} $ in the singlino mass matrix,
while the bino and winos are lighter than the Higgsinos and singlinos if the gluino mass is not far above the current lower bound at the LHC around 1.3 TeV \cite{Chatrchyan:2013wxa, Aad:2013wta}. 
However, for the model (\ref{eq:model2}) of small $ \kappa_1 $ with general gauge mediation, 
$ m_{S_1}, \mu_2' $ and $ \mu_\textrm{eff} $ can be much smaller than the typical SUSY scale $\msusy$ with $ \mu_2' < \mu_\textrm{eff} \lesssim \sqrt{D_A} < \msusy $ as discussed before. In this case, there can be large mixing between the Higgsinos and singlinos, 
and they can be lighter than the gauginos. 
Also notice that the singlinos are almost Dirac-like when $\kappa_1$ is small, because $ \mu_2' \ll m_{S_1} $. 

\section{Conclusion}\label{Sec:Conclusion}

 In this paper, we have studied singlet-extended SUSY models in the presence of a PQ symmetry, which originates from an anomalous $U(1)_A$ gauge symmetry, to realize a TeV scale SUSY scenario with less than a percent level fine-tuning as allowed by the current experimental bounds.
 An anomalous $U(1)_A$ symmetry broken by the St\"uckelberg mechanism provides not only a plausible origin of a PQ symmetry, but also soft scalar masses through the $D$-term mediation.
 Especially, we consider the specific case that the SUSY breaking modulus takes the no-scale form at leading order, and the $U(1)_A$ and visible sectors are sequestered from the SUSY breaking modulus by one-loop order.
 As a result, the anomaly mediation is negligible, while the moduli mediation is one-loop suppressed compared to the $D$-term mediation.
 Moreover, a spontaneous PQ breaking at the intermediate scale, induced by tachyonic soft scalar mass coming from $D$-term mediation, also results in a spontaneous SUSY breaking with a hierarchical VEV structure by one-loop factor as in Eq. (\ref{Eq:PQhier}).
 This SUSY breaking in the PQ breaking sector is transmitted to the MSSM sector through the gauge mediation, whose magnitude is comparable to the $D$-term mediation.  Consequently, superparticle masses around TeV scale without the SUSY flavor problem can be realized through 
 the mixed $D$-term and gauge mediation.

We have examined the implication of this UV setup to low energy physics around TeV scale. 
To explain the observed Higgs boson mass with TeV scale superpartners, the general singlet-extended 
Higgs sector is considered for the low energy models. We find that at least two singlet fields are necessary to complete a low fine-tuned SUSY
scenario with a relatively simple PQ breaking sector, and the forms of models are quite constrained by our UV setup so that 
there are only two working models with two singlet fields.

Some of the phenomenological consequences of the two minimal models are investigated. For the model (\ref{eq:model1}),  
the singlet Higgs sector must be as heavy as other superparticles around $\msusy \sim 1$ TeV scale for a consistent EWSB. 
On the other hand, the singlet sector can be parametrically lighter than the other sectors for the other model (\ref{eq:model2}) with the dimensionless couplings $\kappa_1$, $\kappa_2$ smaller than order unity, when a general gauge mediation is realized. 
This can lead to significant departure from the SM Higgs boson properties by singlet mixing with testable signatures.
Further phenomenological studies of the models will be done in future works.

\appendix

\section{Estimation of the soft terms in the large volume scenario framework}\label{Sec:appA}
 
 In this appendix, we briefly estimate the soft terms in the large volume scenario (LVS) framework, which provides negligible anomaly mediation and $\epsilon_1 \sim \epsilon_2 \sim 1/8\pi^2$.
 Detailed calculation with an explicit example can be found in Refs. \cite{Choi:2010gm, Shin:2011uk}.
 
  In the LVS, a volume modulus $T_b$ is stabilized such that $t_b\equiv T_b+T_b^* \gg 1$ in the string length unit $M_{\rm string}=1$ to give a large compactification volume.
  This can be achieved by introducing another K\"ahler modulus $T_s$ determining a volume of a small cycle.
  When $T_s$ admits the non-perturbative effect $e^{-aT_s}$ in the superpotential, the $\alpha'$-correction of ${\cal O}(1/t_b^{3/2})$ in the K\"ahler potential competes with the effect so that  $T_b$ is stabilized at an exponentially large value \cite{Balasubramanian:2005zx, Conlon:2005ki}.
  For example, in type IIB theory, Calabi-Yau(CY) three-fold volume is given by ${\cal V}_{\rm CY}\sim t_b^{3/2} \sim |e^{a T_s}|$ and this is just $\mplanck^2/M_{\rm string}^2$.
  
  On the other hand, the MSSM sector and an anomalous $U(1)_A$ sector are supported by the visible sector 4-cycle whose volume is determined by a new modulus $T_A$. 
The modulus $T_A$ cannot be identified with $T_b$ since it gives too small SM gauge couplings $g_{\rm SM}^2 \sim 1/T_b$.
  Moreover, $T_A$ cannot have a D3 instanton superpotential, the essential feature of $T_s$ to stabilize $T_b$ \cite{Blumenhagen:2007sm}, therefore $T_s$ cannot play a role of $T_A$.
  Since an instanton superpotential $e^{-aT_A}$ is absent, the modulus $T_A$ should be stabilized in another way.
  We consider the case where $T_A$ is stabilized through the $D$-term of the anomalous $U(1)_A$ gauge multiplet, which results in the $D$-term mediation.
  
 Now consider the dynamics of $U(1)_A$ and visible sectors.  
 At leading order,   $T_b$ has the no-scale structure satisfying $K=-3\ln t_b$ and $\partial W/\partial T_b=0$.
  Subleading effects would appear as expansions in large volume $1/t_b$ and quantum correction $\alpha_{s, A} \ln t_b$.
  Let us define $\tilde{t}_s \equiv t_s-\alpha_s \ln t_b$ and $\tilde{t}_A \equiv t_A-\alpha_A \ln t_b$ \cite{Conlon:2010ji}.
  Then, the generic forms of  K\"ahler potential, superpotential, and gauge kinetic functions for $U(1)_A$ and MSSM sectors are given by
  \dis{K=&K_0(t_b, t_s, t_A)+Z_i \Phi_i^* e^{2q_i V_A}\Phi_i
  \\
  =&-3\ln t_b + \frac{1}{t_b^p}K_{0,1}(\tilde{t_s})+\frac{1}{t_b^n}\Big[\Omega_0(\tilde{t}_s, \tilde{t}_A)+ \frac{1}{t_b^{p'}}\Omega_1(\tilde{t}_s, \tilde{t}_A ) \Big]
  \\
  &+\frac{1}{t_b}\Big[{\cal Y}_{i,0}(\tilde{t}_s, \tilde{t}_A)+\frac{1}{t_b^{p''}}{\cal Y}_{i,1}(\tilde{t}_s, \tilde{t}_A)\Big]\Phi_i^* e^{2q_i V_A}\Phi_i, \label{Eq:genKahler}}
  \dis{W=W_0(T_s, T_A)+\frac{1}{3!}\lambda_{ijk}(T_s, T_A)\Phi_i\Phi_j\Phi_k+\frac{1}{n!}\kappa_{i_1i_2..i_n}(T_s, T_A)\Phi_{i_1}\Phi_{i_2}\cdots\Phi_{i_n},}
  \dis{f_A=\gamma_A(T_s)+k_A T_A,\quad\quad f_a=\gamma_a(T_s) +k_a T_A}
  where $p, p', p''$ and $n$ are some positive integer, especially $p=3/2$ in type IIB string theory. 
  The matter K\"ahler metric $Z_i$ is in the form of $Z_i=(1/t_b){\cal Y}_i$ such that it does not have power-law dependence on the CY3 volume. 

   More explicitly, we take a K\"ahler potential and superpotential
  \dis{&K=-3\ln t_b +\frac{2(\tilde{t}_s^{3/2}-\xi_{\alpha'})}{t_b^{3/2}}+\frac{1}{2 t_b^p}\big(\tilde{t}_A^2 +{\cal O}(t_A^3)\big)+Z_{X} X^* e^{2 q_1V}X + Z_{Y} Y^* e^{2 q_2V}Y, 
  \\
  &W=W_0 + A e^{-a T_s} + y\frac{X^{n+2}Y}{M_*^n},\label{Eq:LVSexample}}
 as investigated in Ref. \cite{Choi:2010gm}.
 In this example, $t_b$ is stabilized at large value,
 \dis{&t_b^{3/2}=e^{at_2/2}\frac{W_0}{aA}\xi
 _{\alpha'}\Big[\frac32-\frac{21+8a\alpha_s}{12a\tilde{t}_s}+{\cal O}\Big(\frac{1}{(a\tilde{t}_2)^2}\Big)\Big],
 \\
 &\tilde{t}_s^{3/2}=\xi_{\alpha'}\Big[1+\frac{3-13a\alpha_s}{3a\tilde{t}_s}+{\cal O}\Big(\frac{1}{(a\tilde{t}_2)^2}\Big)\Big].}
 Once $t_b$ ans $t_s$ are stabilized, we have the effective potential of the PQ sector fields, $\{t_A, X, Y\}$ as Eq. (A2) of Ref. \cite{Choi:2010gm}.
 As a result, $t_A$ is stabilized as $\tilde{t}_A =\delta_{\rm GS}v_{\rm PQ}^2/M_{\rm GS}^2+{\cal O}(\delta_{\rm GS}^2)$ where $M_{\rm GS}^2$ is calculated to be $(\delta_{\rm GS}/2)^2(1/t_b^p)$, whereas $X, Y$ have the intermediate scale VEVs as discussed in Sec. \ref{subsec:PQbreaking}.
 This example confirms various features used in our setup listed below.

 Due to the no-scale structure at leading order, the anomaly-mediation effect is negligibly small,
  \dis{\frac{F^C}{C}={\cal O}\Big(m_{3/2}\frac{1}{t_b^p}, m_{3/2}\frac{|\phi|^2}{\mplanck^2}\Big).}
  On the other hand, FI term is extremely suppressed,
   \dis{\xi_{\rm FI}\simeq \frac{\delta_{\rm GS}}{t_b^n}\Big(\partial_{t_A}\Omega_0+{\cal O}\Big(\frac{1}{t_b^{p'}}\Big)\Big)={\cal O}\Big(\frac{|\phi|^2}{\mplanck^2 t_b}\Big),}
 where the last relation implies that an almost vanishing $D$-term is a result of cancellation between FI term and matter contribution to $D$-term. 
 This is explicitly checked in the example Eq. (\ref{Eq:LVSexample}) as
 \dis{\xi_{\rm FI}=M_{\rm GS}^2\frac{2\tilde{t}_A}{\delta_{\rm GS}}=2\frac{v_{\rm PQ}^2}{M_{\rm pl}^2}+{\cal O}(\delta_{\rm GS}),}
 When matter VEVs are developed as a result of SUSY breaking, we can say  FI term vanishes in the supersymmetric limit.
 Actually, $D$-term given by
 \dis{g_A^2D_A  \sim \Big(\frac{\partial_{t_A}\partial_{t_b}^2 K_0}{\partial_{t_A}^2 K_0 \partial_{t_b}^2 K_0}\Big)\frac{|m_{3/2}|^2}{\delta_{\rm GS}} \equiv \frac{\epsilon_1}{\delta_{\rm GS}}|m_{3/2}|^2,}
 is estimated as
 \dis{\epsilon_1=\frac{\partial_{t_A}\partial_{t_b}^2 K_0}{\partial_{t_A}^2 K_0 \partial_{t_b}^2 K_0} \sim \frac{1}{\partial_{t_A}^2\Omega_0}(a_1 \partial_{t_A}\Omega_0 + a_2 \alpha_A \partial_{t_A}^2\Omega_0 +a_3 \alpha_A^2 \partial_{t_A}^3\Omega_0 ),}
 where $a_{1,2,3}$ are order one coefficients. 
 Whereas $\partial_{t_A}\Omega_0$ is suppressed due to FI term suppression, $\partial_{t_A}^2\Omega_0$ can be a coefficient of order unity, from {\it e.g.} $\Omega_0 = \tilde{t}_A^2$, so we find that $\epsilon_1 \sim \alpha_A \sim 1/8\pi^2$.
 Finally, using the fact that $t_b$ comes in ${\cal Y}_i$ by $-\alpha_{s, A}\ln t_b$ through the combinations ${\tilde t}_{s, A}$, we obtain
 \dis{\epsilon_2 m_{3/2} \sim  F^{T_b}\partial_{t_b}\ln({\cal Y}_{i,0}) \sim m_{3/2} t_b \frac{\partial_{t_b} {\cal Y}_i}{{\cal Y}_i}  \sim  m_{3/2} \alpha_{s, A}\frac{\partial_{t_{s, A}} {\cal Y}_i}{{\cal Y}_i},}
 and when $\partial_{t_{s, A}} {\cal Y}_i /{\cal Y}_i \sim {\cal O}(1)$, we have $\epsilon_2 \sim \alpha_{s, A} \sim 1/8\pi^2$.

 \section{Minimal low energy models} \label{Sec:appB} 
 
In this appendix, the detailed procedure to obtain the minimal models (\ref{eq:model1}) and (\ref{eq:model2}) is presented.
In Sec. \ref{subsec:Onesinglet}, we will show that one singlet extension with a simple PQ breaking sector is not viable for the natural EWSB.
As the next minimal possibility, two singlets extension is discussed in Sec. \ref{subsec:Twosinglets}, where those two minimal models
are found. In Sec. \ref{subsec:nmPQ}, non-minimal PQ breaking sectors are investigated, by which one singlet extension
can be made viable.

\subsection{One singlet extension with a simple PQ breaking sector}\label{subsec:Onesinglet}
The simplest singlet extended Higgs sector will be just one singlet field extension by the singlet superpotential $ f(S_H, S_j)= f(S_H) $ without any other singlet field $ S_j $ interacting with $ S_H $.
In this case, the general 
form of $ f(S_H) $ is given by
  \dis{f(S_H)=\xi(1 + \theta^2 C)S_H+\frac12 \mu'(1 +\theta^2 B')S_H^2.}
 Notice that $ S_H^3 $ is suppressed by a small coupling less than $ (v_{PQ}/M_*)^p $ as $ S_H $ is charged under a PQ symmetry. 
Solving equations of motion, $ \mu_\textrm{eff} $ and $ (B\mu)_\textrm{eff} $ are found to be
\dis{
&\mu_{\rm eff}=\frac{\lambda}{2}\frac{-2\xi \mu'- 2C \xi +\lambda(A_\lambda+\mu')v^2 \sin 2\beta}{m_S^2+\lambda^2v^2+{\mu'}^2+B'\mu'},
\\
&(B\mu)_\textrm{eff} = \lambda \xi +\mu'\mu_{\rm eff}+A_\lambda\mu_{\rm eff}.
} 
From these equations, one can find that at least two parameters among $( \xi, ~C\xi, ~\mu') $ should be around
$m_{3/2}$ to make $ \mu_\textrm{eff}$ and ${(B\mu)_\textrm{eff}}$ around $\msusy$,
when $ A_\lambda $ is negligible.\footnote{Even though $A$-term by the moduli mediation is suppressed by $\epsilon$ compared to $\msusy$ as explained in Sec. \ref{Sec:Model}, sizable $ A_\lambda$ might be generated by RG running, dominantly from gaugino masses. However, it turns out to be still not large enough to realize 
the desired EWSB.}
However, the sequestering factor $\epsilon \sim 1/8\pi^2$ makes it non-trivial. 
For example, we obtain $ \xi \sim m_{3/2}^2 $ from the following operator,
\dis{
\Delta W = \frac{X^{2n+2}}{M_*^{2n}}S_H \sim m_{3/2}^2 ( 1+ \theta^2 \epsilon m_{3/2}) S_H.
}
 On the other hand, $ C\xi $ is also generated around $ \epsilon m_{3/2}^3 \sim m_{3/2}^3/8\pi^2 $ from the same operator.
Thus we cannot obtain $\xi$ and $C\xi$ of $\Omsusy$ simultaneously from the operator due to the $\epsilon$ factor. This situation
is actually generic for an arbitrary single operator because of the structure  $X\sim v_{PQ}(1 + \theta^2 \epsilon m_{3/2})$, $ Y\sim v_{PQ}(\epsilon + \theta^2 m_{3/2})$.
Hence one cannot make  
two parameters among $( \xi, ~C\xi, ~\mu') $ be around $\msusy$ from a single higher dimensional operator.\footnote{Recall that the factor 
$ \epsilon \sim 1/8\pi^2 $ makes it possible for the gauge mediation to be comparable to $ m_{3/2} $ by $F^{Y}/Y \sim  m_{3/2}/\epsilon$. In this sense, the situation is similar to the $ \mu/B\mu $ problem
in the gauge mediation \cite{Dvali:1996cu}.}

We are thus led to have at least two higher dimensional operators involving the singlet field for the desired EWSB to occur.
However, PQ charges of the three fields ($X, Y, S_H$) are already determined from $ W_{\rm PQ} = X^{n+2}Y/M_*^n $
and a higher dimensional operator for one parameter among $( \xi, ~C\xi, ~\mu') $ to be around $\msusy$. 
It means that we cannot arbitrarily write down another higher dimensional operator for another singlet parameter.

In Table  \ref{Table:1singlet}, we enumerate all possible operators and corresponding PQ charges to 
produce each parameter of $( \xi, ~C\xi, ~\mu') $ around $\msusy$.\footnote{If we allow that the effective suppressing mass scale of higher dimensional operators varies up to one-loop factor over different operators by varying the magnitude of their dimensionless coefficients, there can 
be more possibilities than the cases listed in this table because $ \epsilon $ suppression can be overcome by one loop factor smaller mass scale 
of a higher dimensional operator. We have examined these possibilities also, but it cannot change the conclusions derived here including the cases of more singlet fields in the next subsection.}   
\begin{table}
\begin{center}
\begin{tabular}{| c || c | c | c | c |} 
\hline
  &  $\Delta K$  & $\Delta W$ & $(q_{X}, q_{Y})$ & Dangerous term \\ \hline \hline
{$\xi$} & $\frac{X^n Y^*}{M_*^n}S$ & $\frac{X^{2n+2}}{M_*^{2n}}S$  & $\left(-\frac{1}{2(n+1)}, \frac{n+2}{2(n+1)}\right)q_S $ & \\ \cline{2-5}
&  $\frac{{X^*}^n Y^*}{M_*^n}S $ &   & 
$\left(-\frac{1}{2}, \frac{n+2}{2}\right)q_S$ & $X^2 S$  \\ \hline
$C\xi$ &  & $\frac{X^{2n+1} Y}{M_*^{2n}} S$ & $\left(-\frac{1}{n-1}, \frac{n+2}{n-1}\right)q_S$ & $ X^{n-1} S $ \\ \hline
$\mu'$ &  & $\frac{X^{n+1}}{M_*^{n}} S^2$ & $\left(-\frac{2}{n+1}, \frac{2(n+2)}{n+1}\right)q_S$ & \\ \hline
\end{tabular}  
\end{center}
\caption{All higher dimensional operators and PQ charge assignments to give each of $ \xi, ~C\xi,~\mu' $ around $m_{3/2}$. For some cases, there appear unavoidable
dangerous tadpoles.
\label{Table:1singlet}}
\end{table}   
In the table, one can see that there is no PQ charge assignment that can make two parameters among $( \xi, ~C\xi, ~\mu') $ be simultaneously around $\msusy$.
Moreover, for some cases, there appear unavoidable dangerous
tadpoles which produce $ \mu_\textrm{eff} $ too larger than $\msusy$.
 These tadpoles cannot be forbidden even if one imposes additional symmetries, because 
they must be allowed as long as the operators in the table and $ W_{\rm PQ} = X^{n+2}Y/M_*^n $ are allowed by symmetries.
This table will turn out to be useful in the following subsections also.  
Therefore, we conclude that one singlet extension with the simple PQ breaking sector given in Sec. \ref{subsec:PQbreaking} is not viable for the low fine-tuned
EWSB.



\subsection{Two singlets extension}\label{subsec:Twosinglets}
 The next minimal possibility will be two singlet extended Higgs sectors with $ f(S_H, S_j)=f(S_H, S_1) $. A generic form of 
 $f(S_H, S_1)$ is
\dis{f(S_H, S_1) &=\frac12\kappa(1+\theta^2 A_\kappa) S_H^2S_1+\frac12 \kappa_1 (1+\theta^2 A_{\kappa_1}) S_HS_1^2 +M_1(1 + \theta^2 B_1)S_H S_1
  \\
  +&~\xi(1 + \theta^2 C)S_H+\frac12 \mu'(1 + \theta^2 B') S_H^2 +\xi_1(1 + \theta^2 C_1)S_1
  +\frac12 \mu_1' (1 + \theta^2 B_1')S_1^2, \label{Eq:twose}} 
where $A$-terms are suppressed by $ \epsilon $ compared to $\msusy$, so we will neglect $ A_\kappa$ and $A_{\kappa_1} $. 

Non-zero PQ charges of the fields $(X, Y, S_H, S_1)$ will be fixed by three operators.
Thus besides the operator 
$ W_{\rm PQ} = X^{n+2}Y/M_*^n $, we can write down two more operators involving the singlet fields $S_H, S_1$ as desired forms.  
Note that working models must involve at least one interaction term
 between $ S_H $ and $ S_1 $, like $ S_H^2 S_1,~ S_H S_1^2 $, or $S_H S_1$,   
since otherwise the situation is not actually different from the one singlet extension of the previous subsection, which is shown to be unviable. 
Therefore, we should find 
the models in which \textit{two terms} in (\ref{Eq:twose}) including one of the interaction terms yield
$ \mu_\textrm{eff} \sim \langle S_H \rangle \lesssim \Omsusy$ and
$(B\mu)_\textrm{eff} \sim \langle \partial_{S_H} f \rangle \sim \Omsusysq $.
 
There are two ways for $ S_H  $ to get a non-zero VEV. With the $ \kappa S_H^2 S_1 $ term, we get a quartic scalar potential
 for $ S_H $ so that it can get a non-zero VEV by its tachyonic mass.
 In the absence of this term, one can see from the general form (\ref{Eq:twose}) that the scalar potential for $ S_H $ can be quadratic at most, so a tadpole scalar potential of $ S_H $ can make a non-zero VEV for $S_H$ if $S_H$ is non-tachyonic.
Thus we will investigate models with/without the $ \kappa S_H^2 S_1 $ term in the following.
 
\subsubsection{With the cubic interaction $ S_H^2 S_1 $ in $ f(S_H, S_1) $}

The PQ symmetric cubic interaction term $ \kappa S_H^2 S_1 /2 $ in $ f(S_H, S_1) $ gives a quartic scalar potential $  \kappa^2 |S_H|^4/4 $, so $ S_H $ can 
get its VEV by $ |S_H| \sim |m_{S_H}|/\kappa $ if it is tachyonic 
($ m_{S_H}^2 < 0  $). Note that the singlet fields get their masses mainly from
the $D$-term mediation with subdominant RG running effect.
Therefore, if $ S_H $ is tachyonic, then $ S_1 $ must be non-tachyonic because the sign of their soft mass squared is determined by their PQ charges in the $D$-term mediation.
In this case, the right size of $(B\mu)_\textrm{eff} \sim \partial_{S_H} f \sim \kappa S_H S_1 + \cdots$ is to be obtained either by 
$  \langle S_1 \rangle \sim \msusy $ or by another supersymmetric term in $ f(S_H, S_1) $ when $ \langle S_1 \rangle $ vanishes. 

First, let us consider the way to obtain $(B\mu)_\textrm{eff} $ by
non-vanishing $ \langle S_1 \rangle$. Since $ S_1 $ is not tachyonic, it needs a tadpole or cubic scalar potential for a non-zero VEV.
A cubic scalar potential for $ S_1 $ can be obtained from another superpotential term
$ \kappa_1 S_H S_1^2 $, but this term is not allowed by the PQ symmetry unless $ S_H$ and $S_1 $ are uncharged, because of $ \kappa S_H^2 S_1 $ term.
 On the other hand, in order to obtain a tadpole scalar potential for $ S_1 $, one can find that there are five possible models :
\dis{ \label{S1tad}
f(S_H, S_1) = \frac12 \kappa S_H^2 S_1 + \left(\theta^2 C_1\xi_1 S_1, ~
\theta^2 B_1M_1 S_H S_1, ~ \xi S_H, ~ \frac12 \mu' S_H^2, ~\textrm{or}~~ \frac12 \mu_1' S_1^2\right).
}  
The first model $ \theta^2 C_1 \xi_1 S_1 $ is not viable, because it allows unavoidable large tadpole superpotential $ X^{n-1} S_1 $ as shown in Table \ref{Table:1singlet} so that it induces 
$(B\mu)_\textrm{eff} \gg \Omsusysq $. The second model $ \theta^2 B_1M_1 S_H S_1 $ can be shown to be only realized by the superpotential term 
$ X^n Y S_1 S_H / M_* $, but this case also suffers from a large tadpole $ X^2 S_H $ which results in $ \mu_\textrm{eff} \gg \Omsusy $.
The third and fourth models $ \xi S_H, \mu' S_H^2/2 $ require
such a PQ charge assignment for $ S_H $ that $ S_H $ is non-tachyonic in order to render $ X $ tachyonic for a spontaneous
PQ symmetry breaking, as seen in Table \ref{Table:1singlet}. Hence it will make $ S_1 $ tachyonic, and then it can be shown that $ S_1 $ is destabilized and $ \langle S_H \rangle $ vanishes to make $ \mu_\textrm{eff} = 0 $, so they are excluded.
Finally, the last model turns out to give a consistent scenario
with $ \mu_1' \sim X^{n+1}/M_*^n  $ which allows non-tachyonic $ S_1 $ 
and tachyonic $ S_H $. 
Therefore we have found a working model :
\dis{ \label{viable_model1}
f(S_H, S_1) = \frac12 \kappa_1 S_H^2 S_1 + \frac12 \kappa_2 \frac{X^{n+1}}{M_*^n} S_1^2.
}

The other way to obtain $(B\mu)_\textrm{eff} $ is through another supersymmetric term 
besides $ \kappa S_H^2 S_1 $ in $ f(S_H, S_1) $ when $ \langle S_1 \rangle $
vanishes. This could be achieved by either $ \xi S_H $ or $ \mu' S_H^2/2 $ giving $ \partial_{S_H}f \sim (\xi, ~\mu' S_H) \sim  \Omsusysq $ if $ S_H $ is consistently stabilized through its quartic scalar potential. However, these cases are already included in (\ref{S1tad}), which turned out to be 
not working.

 Finally, if $ S_H $ is not tachyonic, there should be a tadpole or cubic scalar potential for $S_H$ to get a non-zero VEV even with its quartic scalar potential.
Barring the cases already included in (\ref{S1tad}), we find that the following
models can give such a scalar potential for $ S_H $,
\dis{
f(S_H, S_1) = \frac12 \kappa S_H^2 S_1 + \left( \theta^2 C\xi S_H, ~\textrm{or} ~~ M_1 S_H S_1\right).
}
The first case $ \theta^2 C\xi S_H $ suffers from the large
tadpole superpotential $ X^{n-1} S_H $ as discussed before, and then $ (B\mu)_\textrm{eff} \sim \partial_{S_H} f \sim M_*^{3-n} X^{n-1} $ is much larger than $ \Omsusysq $.
The second case $ M_1 S_H S_1 $ can be only realized by $ M_1 \sim X^{n+1}/M_*^n $. However, this also allows an unavoidable tadpole superpotential $ XYS_H $. Therefore, we conclude that there is only one 
working model with the cubic interaction $ S_H^2 S_1 $ in $ f(S_H, S_1) $. 

\subsubsection{Without the cubic interaction $ S_H^2 S_1 $ in $ f(S_H, S_1) $}

In the absence of the cubic interaction $ S_H^2 S_1 $ in $ f(S_H, S_1) $, 
the scalar potential for $ S_H $ can be quadratic at most. Thus $ S_H $ needs
a tadpole scalar potential with its non-negative mass squared to get a non-zero VEV.  
The first obvious choice is through soft terms which are linear in $ S_H $,
\dis{
f(S_H, S_1) = \left( \theta^2 C\xi S_H, ~\textrm{or}~~\theta^2 B_1M_1 S_1 S_H \right) + (\textrm{something}).
}
The $ \theta^2 C\xi S_H $ term is not available because of the large tadpole problem as in the previous cases. The second case $ \theta^2 B_1M_1 S_1 S_H  $ requires that $ S_1 $ gets a non-zero VEV to give a tadpole for $S_H$, and this should be done by another term.
That term must be a supersymmetric term to give non-negligible 
$ (B\mu)_\textrm{eff} \sim \partial_{S_H} f $.
Then one can find that there is only one possibility, which is through $ \kappa_1 S_1^2 S_H/2 $ with tachyonic $ S_1 $.  However, this is similar to the second 
case of (\ref{S1tad}), so it suffers from a large tadpole $ X^2 S_1 $ which produces $ (B\mu)_\textrm{eff} \gg \Omsusysq $.
 
The second way to get a tadpole scalar potential for $ S_H $ is through supersymmetric terms. Such models are found to be
 \begin{eqnarray}
 f(S_H, S_1) & = & M_1 S_1 S_H + \left( \xi_1 S_1, ~\textrm{or}~~\frac12 \mu_1' S_1^2 \right), \label{SHtad1} \\
  f(S_H, S_1) & = & \frac12 \kappa _1 S_1^2 S_H + \left( \xi_1 S_1, ~\frac12 \mu_1' S_1^2,~\textrm{or}~~\frac12 \mu' S_H^2 \right). \label{SHtad2}  
\end{eqnarray}
In the first model $ \xi_1 S_1 $ in (\ref{SHtad1}), there is no scalar potential for $ S_1 $ other than its mass term so that $ (B\mu)_\textrm{eff} \sim \partial_{S_H}f  \sim M_1 S_1 $ is destabilized or vanishes depending on whether $S_1$ is tachyonic or non-tachyonic. The second model
  $ \mu_1' S_1^2/2 $ in (\ref{SHtad1}) has only bilinear scalar potentials for $ S_H $ and $ S_1 $, so they will either have a vanishing VEV, or be destabilized, meaning 
both $ \mu_\textrm{eff}  $ and $ (B\mu)_\textrm{eff} $ cannot have the right size. 

The first and second models of (\ref{SHtad2}) are similar to the third and 
fourth models of (\ref{S1tad}) by just interchaging $ S_H \leftrightarrow S_1 $, and so now $ S_H $ is destabilized and $ \langle S_1 \rangle $ vanishes.
Thus they result in $ \mu_\textrm{eff} \gg \Omsusy $ and 
 $ (B\mu)_\textrm{eff} \sim 0 $. The last case of (\ref{SHtad2}) is also similar to the working model of (\ref{viable_model1}) by interchaging $ S_H \leftrightarrow S_1 $. The only difference is that now $ S_1 $ gets a non-zero VEV by
its quartic scalar potential with a tachyonic mass and $ S_H $ through its tadpole scalar potential after $ \langle S_1 \rangle$ becomes non-zero. Therefore, we conclude that there is one more viable model without 
 the cubic interaction $ S_H^2 S_1 $ in $ f(S_H, S_1) $ : 
 \dis{ \label{viable_model2}
f(S_H, S_1) = \frac12 \kappa_1 S_1^2 S_H + \frac12 \kappa_2 \frac{X^{n+1}}{M_*^n} S_H^2.
}

\subsection{Non-minimal PQ breaking sector} \label{subsec:nmPQ} 
In Sec. \ref{subsec:Onesinglet}, we see that the one singlet extension with a simple PQ breaking sector is not working because the number of fields
is not enough so that PQ charges of each field is determined by smaller number of operators than necessary.
Thus if the spontaneous PQ breaking sector consists of more than two fields, the restriction can be removed. In this subsection, 
we will explore the possibility that the one singlet extension can be made viable with more than two fields in the spontaneous PQ breaking sector.

Let us start with three fields $X, Y, Z$.
One of the three fields must be tachyonic if they are not PQ singlet, because at least
one should have a positive PQ charge to conserve the PQ symmetry. We will call this field $ X $ as in the case of two fields in Sec. \ref{subsec:PQbreaking}. Likewise, at least
one of them should have a negative PQ charge so that it is non-tachyonic,
and we call this field $ Y $. 
The superpotential for the PQ breaking sector must consist of only one operator in order to allow two higher dimensional operators involving $S_H$
to be in desired forms. 
Therefore, the most
general form of the PQ breaking sector with three fields can be written as
\dis{
W_{\rm PQ} = \frac{X^{n_1} Y^{n_2} Z^{n_3}}{M_*^n},
}
where $ n_1 + n_2 + n_3 = n+3 $ and $ n, n_i \geq 1 $. The corresponding scalar potential containing soft terms is
\dis{ \label{Eq:VPQ_3fields}
V_{\rm PQ} = &m_{X}^2 |X|^2 + m_{Y}^2 |Y|^2 + m_{Z}^2 |Z|^2 + A_3\frac{X^{n_1} Y^{n_2} Z^{n_3}}{M_*^n} \\
& +\frac{|X|^{2(n_1-1)} |Y|^{2n_2}
|Z|^{2n_3}}{M_*^{2n}}
+ \frac{|X|^{2n_1} |Y|^{2(n_2-1)}
|Z|^{2n_3}}{M_*^{2n}} + \frac{|X|^{2n_1} |Y|^{2n_2}
|Z|^{2(n_3-1)}}{M_*^{2n}}, 
}
with $ m_{X}^2 < 0$ and $m_{Y}^2 > 0  $ in our convention.

Now we require that at least one non-tachyonic state gets a relatively small VEV proportional to the small $ A_3 $ to generate a sizable gauge mediation as discussed in Sec. \ref{subsec:PQbreaking}. To this end, one can find that $ n_2 $  should be 1 so that the $A_3$-term is linear in $Y$.
Also we observe that $ X, Z $ must get their non-zero VEVs when $Y=0$ in order to make $A_3$-proportional tadpole term for $Y$.
Then $Y$ can get its small VEV proportional to $A_3$. 

With $ Y=0 $, we examine the potential in arbitrary field directions in $X$-$Z$ plane by parametrizing the fields as $ |X|=|\varphi| \cos \alpha, |Z|=|\varphi| \sin \alpha $ with $ 0 \leq \alpha \leq \pi/2$.
In the field direction $\varphi$ with a constant value of $ \alpha $, the potential becomes
\dis{
V_{\rm PQ} = (m_{X}^2 \cos^2 \alpha + m_{Z}^2 \sin^2 \alpha)|\varphi|^2 + (\cos\alpha)^{2n_1} (\sin\alpha)^{2n_3} \frac{|\varphi|^{2(n+2)}}{M_*^{2n}},
}     
where $ m_{X}^2 < 0 $. If $ (m_{X}^2 \cos^2 \alpha + m_{Z}^2 \sin^2 \alpha) < 0 $, the potential will be minimized along $\varphi$ direction at
\dis{
 |\varphi| &= \frac{1}{(\cos\alpha)^{n_1/(n+1)} (\sin\alpha)^{n_3/(n+1)}} \left( M_*^n \sqrt{\left| m_{X}^2 \cos^2 \alpha + m_{Z}^2 \sin^2 \alpha \right|} \right)^{1/(n+1)}\\
 &\sim \frac{v_{PQ}}{(\cos\alpha)^{n_1/(n+1)} (\sin\alpha)^{n_3/(n+1)}}.
 } 
At this field value, the potential turns out that $V_{\rm PQ} \sim -(\cos \alpha)^{-2n_1/(n+1)} (\sin \alpha)^{-2n_3/(n+1)}$ and $\partial^2V_{\rm PQ}/\partial \alpha^2 < 0  $. 
 Thus this point actually corresponds to a saddle point in the  $X$-$Z$ field space. Also the potential is unbounded from below 
in the field direction $ Z=0~(\alpha=0)$ unless $ n_3=0 $. Therefore, $ \varphi $ is destabilized in the $ X$-$Z $ plane. It means that generically one cannot make the required pattern of PQ symmetry breaking with three fields.
 
We are thus led to consider even more than three fields. 
The simplest possibility with four fields will be
\dis{ \label{4PQbreaking}
W_{\rm PQ} = y_1\frac{X^{n+2} Y}{M_*^n} + y_2\frac{X'^{n+2} Y'}{M_*^n},  
}
where $ (X', Y') $ fields have some different PQ charges from the $ (X, Y) $ fields so that any interaction between them is suppressed. Then each term will realize the correct PQ symmetry breaking pattern as in the two fields case, and we can use the two kinds of fields of different PQ charges to generate two parameters among ($ \xi $, $ C\xi $, or $\mu' $) with the PQ charge assignments in Table \ref{Table:1singlet}. In this way, one can realize the low fine-tuned EWSB with one singlet field extended Higgs sector, but 
it requires such a complication of the PQ breaking sector that there must be more than three fields and a non-trivial PQ charge relation among them. 

\acknowledgments

 We thank Kiwoon Choi for suggestion and participation of this work from beginning to the last stage. 
 We also thank Kwang Sik Jeong and Chang Sub Shin for reading off the draft and giving numerous valuable comments.
 This work was supported by IBS-R018-D1.  




\begin{thebibliography}{99}

\bibitem{SUSYreview1} 
  H.~P.~Nilles,
  ``Supersymmetry, Supergravity and Particle Physics,''  
  Phys.\ Rept.\  {\bf 110}, 1 (1984).  
\bibitem{SUSYreview2} 
  H.~E.~Haber and G.~L.~Kane,
  ``The Search for Supersymmetry: Probing Physics Beyond the Standard Model,''  
  Phys.\ Rept.\  {\bf 117}, 75 (1985).  

\bibitem{Axionreview1}
  J.~E.~Kim,
  ``Light Pseudoscalars, Particle Physics and Cosmology,''
  Phys.\ Rept.\  {\bf 150} (1987) 1.
 \bibitem{Axionreview2}
  J.~E.~Kim and G.~Carosi,
  ``Axions and the Strong CP Problem,''
  Rev.\ Mod.\ Phys.\  {\bf 82}, 557 (2010)
  [arXiv:0807.3125 [hep-ph]].


\bibitem{Murayama:1992dj} 
  H.~Murayama, H.~Suzuki and T.~Yanagida,
  ``Radiative breaking of Peccei-Quinn symmetry at the intermediate mass scale,''
  Phys.\ Lett.\ B {\bf 291}, 418 (1992).
  
\bibitem{PecceiQuinn} 
  R.~D.~Peccei and H.~R.~Quinn,
  ``CP Conservation in the Presence of Instantons,''  
Phys.\ Rev.\ Lett.\  {\bf 38}, 1440 (1977).  
\bibitem{Peccei:1977ur} 
  R.~D.~Peccei and H.~R.~Quinn,
  ``Constraints Imposed by CP Conservation in the Presence of Instantons,''  
Phys.\ Rev.\ D {\bf 16}, 1791 (1977).  


\bibitem{Kim:1983dt} 
  J.~E.~Kim and H.~P.~Nilles,
  ``The mu Problem and the Strong CP Problem,''  
  Phys.\ Lett.\ B {\bf 138}, 150 (1984).  
   
  
\bibitem{Lyth:1995ka} 
  D.~H.~Lyth and E.~D.~Stewart,
  ``Thermal inflation and the moduli problem,''  
  Phys.\ Rev.\ D {\bf 53}, 1784 (1996)  [hep-ph/9510204].  
\bibitem{Choi:1996vz} 
  K.~Choi, E.~J.~Chun and J.~E.~Kim,
  ``Cosmological implications of radiatively generated axion scale,''  
  Phys.\ Lett.\ B {\bf 403}, 209 (1997)  [hep-ph/9608222].  

\bibitem{Papucci:2011wy} 
  M.~Papucci, J.~T.~Ruderman and A.~Weiler,
  ``Natural SUSY Endures,''
  JHEP {\bf 1209}, 035 (2012)
  [arXiv:1110.6926 [hep-ph]].
  

\bibitem{Hall:2011aa} 
  L.~J.~Hall, D.~Pinner and J.~T.~Ruderman,
  ``A Natural SUSY Higgs Near 126 GeV,''  
  JHEP {\bf 1204}, 131 (2012)  [arXiv:1112.2703 [hep-ph]].  
  
\bibitem{Feng:2013tvd} 
  J.~L.~Feng, P.~Kant, S.~Profumo and D.~Sanford,
  ``Three-Loop Corrections to the Higgs Boson Mass and Implications for Supersymmetry at the LHC,'' 
   Phys.\ Rev.\ Lett.\  {\bf 111}, 131802 (2013)  [arXiv:1306.2318 [hep-ph]].  

\bibitem{NMSSM} 
  M.~Maniatis,
  ``The Next-to-Minimal Supersymmetric extension of the Standard Model reviewed,''
  Int.\ J.\ Mod.\ Phys.\ A {\bf 25}, 3505 (2010)
  [arXiv:0906.0777 [hep-ph]].
  
\bibitem{Ellwanger:2009dp} 
  U.~Ellwanger, C.~Hugonie and A.~M.~Teixeira,
  ``The Next-to-Minimal Supersymmetric Standard Model,''
  Phys.\ Rept.\  {\bf 496}, 1 (2010)
  [arXiv:0910.1785 [hep-ph]].

\bibitem{Barbieri:2006bg} 
  R.~Barbieri, L.~J.~Hall, Y.~Nomura and V.~S.~Rychkov,
  ``Supersymmetry without a Light Higgs Boson,'' 
   Phys.\ Rev.\ D {\bf 75}, 035007 (2007)  [hep-ph/0607332].  

\bibitem{Jeong:2011jk} 
  K.~S.~Jeong, Y.~Shoji and M.~Yamaguchi,
  ``Peccei-Quinn invariant extension of the NMSSM,''
  JHEP {\bf 1204}, 022 (2012)
  [arXiv:1112.1014 [hep-ph]].
 
\bibitem{Jeong:2012ma}  
  K.~S.~Jeong, Y.~Shoji and M.~Yamaguchi,
  ``Singlet-Doublet Higgs Mixing and Its Implications on the Higgs mass in the PQ-NMSSM,''
  JHEP {\bf 1209}, 007 (2012)
  [arXiv:1205.2486 [hep-ph]].
  
\bibitem{Kim:2012az} 
  J.~E.~Kim, H.~P.~Nilles and M.~S.~Seo,
  ``Singlet Superfield Extension of the Minimal Supersymmetric Standard model with Peccei-Quinn symmetry and a Light Pseudoscalar Higgs Boson at the LHC,''
  Mod.\ Phys.\ Lett.\ A {\bf 27}, 1250166 (2012)
  [arXiv:1201.6547 [hep-ph]].

\bibitem{Bae:2012am}
  K.~J.~Bae, K.~Choi, E.~J.~Chun, S.~H.~Im, C.~B.~Park and C.~S.~Shin,
  ``Peccei-Quinn NMSSM in the light of 125 GeV Higgs,''
  JHEP {\bf 1211} (2012) 118
  [arXiv:1208.2555 [hep-ph]].
  
\bibitem{Choi:2013lda}
  K.~Choi, S.~H.~Im, K.~S.~Jeong and M.~S.~Seo,
  ``Higgs phenomenology in the Peccei-Quinn invariant NMSSM,''
  JHEP {\bf 1401}, 072 (2014)
  [arXiv:1308.4447 [hep-ph]].


\bibitem{Gherghetta:1995jx}
  T.~Gherghetta and G.~L.~Kane,
  ``Chaotic inflation and a radiatively generated intermediate scale in the supersymmetric standard model,''  
  Phys.\ Lett.\ B {\bf 354}, 300 (1995)  [hep-ph/9504420].  
  
\bibitem{Bae:2014yta}   
  K.~J.~Bae, H.~Baer and H.~Serce,
  ``A natural Little Hierarchy for SUSY from radiative breaking of PQ symmetry,''  
  arXiv:1410.7500 [hep-ph].  


\bibitem{Abbott:1989jw} 
  L.~F.~Abbott and M.~B.~Wise,
  ``Wormholes and Global Symmetries,''  
  Nucl.\ Phys.\ B {\bf 325}, 687 (1989).  
  
\bibitem{Coleman:1989zu} 
  S.~R.~Coleman and K.~-M.~Lee,
  ``Wormholes Made Without Massless Matter Fields,''  
  Nucl.\ Phys.\ B {\bf 329}, 387 (1990).  
  
\bibitem{Kallosh:1995hi} 
  R.~Kallosh, A.~D.~Linde, D.~A.~Linde and L.~Susskind,
  ``Gravity and global symmetries,''  
  Phys.\ Rev.\ D {\bf 52}, 912 (1995)  [hep-th/9502069].  
  
\bibitem{Banks:2010zn} 
  T.~Banks and N.~Seiberg,
  ``Symmetries and Strings in Field Theory and Gravity,''  
  Phys.\ Rev.\ D {\bf 83}, 084019 (2011)  [arXiv:1011.5120 [hep-th]].  

\bibitem{Barr:1992qq} 
  S.~M.~Barr and D.~Seckel,
  ``Planck scale corrections to axion models,''  
  Phys.\ Rev.\ D {\bf 46}, 539 (1992).  
  
\bibitem{Kamionkowski:1992mf} 
  M.~Kamionkowski and J.~March-Russell,
  ``Planck scale physics and the Peccei-Quinn mechanism,''  
  Phys.\ Lett.\ B {\bf 282}, 137 (1992)  [hep-th/9202003].  
  
\bibitem{Holman:1992us} 
  R.~Holman, S.~D.~H.~Hsu, T.~W.~Kephart, E.~W.~Kolb, R.~Watkins and L.~M.~Widrow,
  ``Solutions to the strong CP problem in a world with gravity,''  
  Phys.\ Lett.\ B {\bf 282}, 132 (1992)  [hep-ph/9203206].  

\bibitem{Barr:1985hk} 
  S.~M.~Barr,
  ``Harmless Axions in Superstring Theories,''  
  Phys.\ Lett.\ B {\bf 158}, 397 (1985).  
  
\bibitem{Kim:1988dd} 
  J.~E.~Kim,
  ``The Strong {CP} Problem in Orbifold Compactifications and an SU(3) X SU(2) X U(1)-$n$ Model,''  
  Phys.\ Lett.\ B {\bf 207}, 434 (1988).  

\bibitem{Svrcek:2006yi} 
  P.~Svrcek and E.~Witten,
  ``Axions In String Theory,''  
JHEP {\bf 0606}, 051 (2006)  [hep-th/0605206].  

\bibitem{Choi:2011xt} 
  K.~Choi, K.~S.~Jeong, K.~-I.~Okumura and M.~Yamaguchi,
  ``Mixed Mediation of Supersymmetry Breaking with Anomalous U(1) Gauge Symmetry,'' 
   JHEP {\bf 1106}, 049 (2011)  [arXiv:1104.3274 [hep-ph]].  
   
\bibitem{Honecker:2013mya} 
  G.~Honecker and W.~Staessens,
  ``On axionic dark matter in Type IIA string theory,''  
  Fortsch.\ Phys.\  {\bf 62}, 115 (2014)  [arXiv:1312.4517 [hep-th]].  
  
\bibitem{Choi:2014uaa} 
  K.~Choi, K.~S.~Jeong and M.~S.~Seo,
  ``String theoretic QCD axions in the light of PLANCK and BICEP2,''
  JHEP {\bf 1407}, 092 (2014)
  [arXiv:1404.3880 [hep-th]].

\bibitem{Green:1984sg} 
  M.~B.~Green and J.~H.~Schwarz,
  ``Anomaly Cancellation in Supersymmetric D=10 Gauge Theory and Superstring Theory,''  
  Phys.\ Lett.\ B {\bf 149}, 117 (1984).  

\bibitem{Ibanez:2012zz}
  L.~E.~Ibanez and A.~M.~Uranga,
  ``String theory and particle physics: An introduction to string phenomenology,''  Cambridge, UK: Univ. Pr. (2012) 673 p.  

\bibitem{Cvetic:2001nr} 
  M.~Cvetic, G.~Shiu and A.~M.~Uranga,
  ``Chiral four-dimensional N=1 supersymmetric type 2A orientifolds from intersecting D6 branes,'' 
   Nucl.\ Phys.\ B {\bf 615}, 3 (2001)  [hep-th/0107166].  
   
\bibitem{Cremades:2002te} 
  D.~Cremades, L.~E.~Ibanez and F.~Marchesano,
  ``SUSY quivers, intersecting branes and the modest hierarchy problem,''  
  JHEP {\bf 0207}, 009 (2002)  [hep-th/0201205].  
  
\bibitem{Honecker:2004kb} 
  G.~Honecker and T.~Ott,
  ``Getting just the supersymmetric standard model at intersecting branes on the Z(6) orientifold,''  
  Phys.\ Rev.\ D {\bf 70}, 126010 (2004)  [Erratum-ibid.\ D {\bf 71}, 069902 (2005)]  [hep-th/0404055].  
  
\bibitem{Villadoro:2006ia} 
  G.~Villadoro and F.~Zwirner,
  ``D terms from D-branes, gauge invariance and moduli stabilization in flux compactifications,''  
  JHEP {\bf 0603}, 087 (2006)  [hep-th/0602120].  
  
\bibitem{Gmeiner:2008xq}
  F.~Gmeiner and G.~Honecker,
  ``Millions of Standard Models on Z-prime(6)?,'' 
   JHEP {\bf 0807} (2008) 052  [arXiv:0806.3039 [hep-th]].  

\bibitem{Blumenhagen:2005ga} 
  R.~Blumenhagen, G.~Honecker and T.~Weigand,
  ``Loop-corrected compactifications of the heterotic string with line bundles,'' 
   JHEP {\bf 0506}, 020 (2005)  [hep-th/0504232].  
   
\bibitem{Blumenhagen:2005pm}
  R.~Blumenhagen, G.~Honecker and T.~Weigand,
  ``Supersymmetric (non-)Abelian bundles in the Type I and SO(32) heterotic string,''
   JHEP {\bf 0508} (2005) 009  [hep-th/0507041].  

\bibitem{Anderson:2009nt}
  L.~B.~Anderson, J.~Gray, A.~Lukas and B.~Ovrut,
  ``Stability Walls in Heterotic Theories,''  
  JHEP {\bf 0909} (2009) 026  [arXiv:0905.1748 [hep-th]].  

\bibitem{Choi:2006bh} 
  K.~Choi and K.~S.~Jeong,
  ``Supersymmetry breaking and moduli stabilization with anomalous U(1) gauge symmetry,''
  JHEP {\bf 0608}, 007 (2006)
  [hep-th/0605108].

\bibitem{Kaplunovsky:1993rd} 
  V.~S.~Kaplunovsky and J.~Louis,
  ``Model independent analysis of soft terms in effective supergravity and in string theory,'' 
   Phys.\ Lett.\ B {\bf 306}, 269 (1993)  [hep-th/9303040].  
   
\bibitem{Brignole:1993dj} 
  A.~Brignole, L.~E.~Ibanez and C.~Munoz,
  ``Towards a theory of soft terms for the supersymmetric Standard Model,''  
  Nucl.\ Phys.\ B {\bf 422}, 125 (1994)  [Erratum-ibid.\ B {\bf 436}, 747 (1995)]  [hep-ph/9308271].  

\bibitem{Randall:1998uk} 
  L.~Randall and R.~Sundrum,
  ``Out of this world supersymmetry breaking,''
  Nucl.\ Phys.\ B {\bf 557}, 79 (1999)
  [hep-th/9810155].
  
  \bibitem{Giudice:1998xp}
  G.~F.~Giudice, M.~A.~Luty, H.~Murayama and R.~Rattazzi,
  ``Gaugino mass without singlets,''
  JHEP {\bf 9812}, 027 (1998)
  [hep-ph/9810442].
  
  \bibitem{Bagger:1999rd} 
  J.~A.~Bagger, T.~Moroi and E.~Poppitz,
  ``Anomaly mediation in supergravity theories,''
  JHEP {\bf 0004}, 009 (2000)
  [hep-th/9911029].




\bibitem{Kawamura:1996wn} 
  Y.~Kawamura and T.~Kobayashi,
  ``Soft scalar masses in string models with anomalous U(1) symmetry,'' 
   Phys.\ Lett.\ B {\bf 375}, 141 (1996)  [Erratum-ibid.\ B {\bf 388}, 867 (1996)]  [hep-ph/9601365].  
   
   
  \bibitem{Binetruy:1996uv}
  P.~Binetruy and E.~Dudas,
 ``Gaugino condensation and the anomalous U(1),''  
  Phys.\ Lett.\ B {\bf 389}, 503 (1996)  [hep-th/9607172].  
  
  \bibitem{Dvali:1996rj}
  G.~R.~Dvali and A.~Pomarol,
  ``Anomalous U(1) as a mediator of supersymmetry breaking,''  
  Phys.\ Rev.\ Lett.\  {\bf 77}, 3728 (1996)  [hep-ph/9607383].  
  
\bibitem{Kawamura:1996bd}
  Y.~Kawamura and T.~Kobayashi,
  ``Generic formula of soft scalar masses in string models,''
  Phys.\ Rev.\ D {\bf 56} (1997) 3844
  [hep-ph/9608233].
  
  \bibitem{Dvali:1997sr}
  G.~R.~Dvali and A.~Pomarol,
  ``Anomalous U(1), gauge mediated supersymmetry breaking and Higgs as pseudoGoldstone bosons,''
  Nucl.\ Phys.\ B {\bf 522} (1998) 3
  [hep-ph/9708364].

\bibitem{Murakami:2001hk} 
  B.~Murakami, K.~Tobe and J.~D.~Wells,
  ``D term challenges for supersymmetric gauged Abelian flavor symmetries,''
  Phys.\ Lett.\ B {\bf 526}, 157 (2002)
  [hep-ph/0111003].
  
\bibitem{Higaki:2003ig}
  T.~Higaki, Y.~Kawamura, T.~Kobayashi and H.~Nakano,
  ``Anomalous U(1) D term contribution in type I string models,''  
  Phys.\ Rev.\ D {\bf 69}, 086004 (2004)  [hep-ph/0308110].  
  
\bibitem{Kors:2004hz} 
  B.~Kors and P.~Nath,
  ``Hierarchically split supersymmetry with Fayet-Iliopoulos $D$-terms in string theory,''
  Nucl.\ Phys.\ B {\bf 711}, 112 (2005)
  [hep-th/0411201].
   
\bibitem{Babu:2005ui}
  K.~S.~Babu, T.~Enkhbat and B.~Mukhopadhyaya,
  ``Split supersymmetry from anomalous U(1),''
  Nucl.\ Phys.\ B {\bf 720}, 47 (2005)
  [hep-ph/0501079].
  
\bibitem{Dudas:2005vv} 
  E.~Dudas and S.~K.~Vempati,
  ``Large $D$-terms, hierarchical soft spectra and moduli stabilisation,''
  Nucl.\ Phys.\ B {\bf 727}, 139 (2005)
  [hep-th/0506172].
  
\bibitem{ArkaniHamed:1998nu} 
  N.~Arkani-Hamed, M.~Dine and S.~P.~Martin,
  ``Dynamical supersymmetry breaking in models with a Green-Schwarz mechanism,''
  Phys.\ Lett.\ B {\bf 431}, 329 (1998)
  [hep-ph/9803432].
  
\bibitem{Barreiro:1998nd}
  T.~Barreiro, B.~de Carlos, J.~A.~Casas and J.~M.~Moreno,
  ``Anomalous U(1), gaugino condensation and supergravity,''
  Phys.\ Lett.\ B {\bf 445}, 82 (1998)
  [hep-ph/9808244].

  
  \bibitem{GarciadelMoral:2005js}
  M.~P.~Garcia del Moral,
  ``A New mechanism of Kahler moduli stabilization in type IIB theory,''  
  JHEP {\bf 0604}, 022 (2006)  [hep-th/0506116].  
  
\bibitem{Villadoro:2005yq}
  G.~Villadoro and F.~Zwirner,
  ``De-Sitter vacua via consistent $D$-terms,'' 
   Phys.\ Rev.\ Lett.\  {\bf 95}, 231602 (2005)  [hep-th/0508167].  
   
\bibitem{Parameswaran:2007kf}
  S.~L.~Parameswaran and A.~Westphal,
  ``Consistent de Sitter string vacua from Kahler stabilization and $D$-term uplifting,''
  Fortsch.\ Phys.\  {\bf 55}, 804 (2007)
  [hep-th/0701215].
  
\bibitem{Achucarro:2006zf} 
  A.~Achucarro, B.~de Carlos, J.~A.~Casas and L.~Doplicher,
  ``De Sitter vacua from uplifting $D$-terms in effective supergravities from realistic strings,''  
  JHEP {\bf 0606}, 014 (2006)  [hep-th/0601190].  
  
  
\bibitem{Gallego:2008sv}
  D.~Gallego and M.~Serone,
  ``Moduli Stabilization in non-Supersymmetric Minkowski Vacua with Anomalous U(1) Symmetry,''  
  JHEP {\bf 0808}, 025 (2008)  [arXiv:0807.0190 [hep-th]].  
  
\bibitem{Krippendorf:2009zza}
  S.~Krippendorf and F.~Quevedo,
  ``Metastable SUSY Breaking, de Sitter Moduli Stabilisation and Kahler Moduli Inflation,''  
  JHEP {\bf 0911}, 039 (2009)  [arXiv:0901.0683 [hep-th]].  


\bibitem{Dudas:2007nz} 
  E.~Dudas, Y.~Mambrini, S.~Pokorski and A.~Romagnoni,
  ``Moduli stabilization with Fayet-Iliopoulos uplift,''
  JHEP {\bf 0804}, 015 (2008)
  [arXiv:0711.4934 [hep-th]].
  
\bibitem{Dudas:2008qf}
  E.~Dudas, Y.~Mambrini, S.~Pokorski, A.~Romagnoni and M.~Trapletti,
  ``Gauge versus Gravity mediation in models with anomalous U(1)'s,''
  JHEP {\bf 0903}, 011 (2009)
  [arXiv:0809.5064 [hep-th]].

\bibitem{Burgess:2003ic} 
  C.~P.~Burgess, R.~Kallosh and F.~Quevedo,
  ``De Sitter string vacua from supersymmetric D terms,''
  JHEP {\bf 0310}, 056 (2003)
  [hep-th/0309187].
  
\bibitem{Jockers:2005zy}
  H.~Jockers and J.~Louis,
  ``$D$-terms and $F$-terms from D7-brane fluxes,''
  Nucl.\ Phys.\ B {\bf 718}, 203 (2005)
  [hep-th/0502059].
 


\bibitem{Wells:2003tf}
  J.~D.~Wells,
  ``Implications of supersymmetry breaking with a little hierarchy between gauginos and scalars,''
  [hep-ph/0306127].
  
\bibitem{ArkaniHamed:2004fb}
  N.~Arkani-Hamed and S.~Dimopoulos,
  ``Supersymmetric unification without low energy supersymmetry and signatures for fine-tuning at the LHC,''
  JHEP {\bf 0506}, 073 (2005)
  [hep-th/0405159].
  
 \bibitem{Giudice:2004tc}
  G.~F.~Giudice and A.~Romanino,
  ``Split supersymmetry,''
  Nucl.\ Phys.\ B {\bf 699}, 65 (2004)
  [Erratum-ibid.\ B {\bf 706}, 65 (2005)]
  [hep-ph/0406088].
  
\bibitem{ArkaniHamed:2004yi} 
  N.~Arkani-Hamed, S.~Dimopoulos, G.~F.~Giudice and A.~Romanino,
  ``Aspects of split supersymmetry,''
  Nucl.\ Phys.\ B {\bf 709}, 3 (2005)
  [hep-ph/0409232].

\bibitem{Dine:1981za} 
  M.~Dine, W.~Fischler and M.~Srednicki,
  ``Supersymmetric Technicolor,''
  Nucl.\ Phys.\ B {\bf 189}, 575 (1981).
  
\bibitem{Dimopoulos:1981au}
  S.~Dimopoulos and S.~Raby,
  ``Supercolor,''
  Nucl.\ Phys.\ B {\bf 192}, 353 (1981).
  
\bibitem{Dine:1981gu}
  M.~Dine and W.~Fischler,
  ``A Phenomenological Model of Particle Physics Based on Supersymmetry,''
  Phys.\ Lett.\ B {\bf 110} (1982) 227.
  
\bibitem{Nappi:1982hm}
  C.~R.~Nappi and B.~A.~Ovrut,
  ``Supersymmetric Extension of the SU(3) x SU(2) x U(1) Model,''
  Phys.\ Lett.\ B {\bf 113}, 175 (1982).
  
\bibitem{AlvarezGaume:1981wy} 
  L.~Alvarez-Gaume, M.~Claudson and M.~B.~Wise,
  ``Low-Energy Supersymmetry,''
  Nucl.\ Phys.\ B {\bf 207}, 96 (1982).
  
\bibitem{Dimopoulos:1982gm}
  S.~Dimopoulos and S.~Raby,
  ``Geometric Hierarchy,''
  Nucl.\ Phys.\ B {\bf 219}, 479 (1983).

\bibitem{Dine:1994vc} 
  M.~Dine, A.~E.~Nelson and Y.~Shirman,
  ``Low-energy dynamical supersymmetry breaking simplified,''
  Phys.\ Rev.\ D {\bf 51}, 1362 (1995)
  [hep-ph/9408384].
  
\bibitem{Dine:1995ag} 
  M.~Dine, A.~E.~Nelson, Y.~Nir and Y.~Shirman,
  ``New tools for low-energy dynamical supersymmetry breaking,''
  Phys.\ Rev.\ D {\bf 53}, 2658 (1996)
  [hep-ph/9507378].
  
\bibitem{Giudice:1998bp}
  G.~F.~Giudice and R.~Rattazzi,
  ``Theories with gauge mediated supersymmetry breaking,''
  Phys.\ Rept.\  {\bf 322}, 419 (1999)
  [hep-ph/9801271].



\bibitem{Cremmer:1983bf} 
  E.~Cremmer, S.~Ferrara, C.~Kounnas and D.~V.~Nanopoulos,
  ``Naturally Vanishing Cosmological Constant in N=1 Supergravity,''  
  Phys.\ Lett.\ B {\bf 133}, 61 (1983).  
  
\bibitem{Ellis:1983ei} 
  J.~R.~Ellis, C.~Kounnas and D.~V.~Nanopoulos,
 ``Phenomenological SU(1,1) Supergravity,''  
  Nucl.\ Phys.\ B {\bf 241}, 406 (1984).  
  
\bibitem{Ellis:1983sf} 
  J.~R.~Ellis, A.~B.~Lahanas, D.~V.~Nanopoulos and K.~Tamvakis,
  ``No-Scale Supersymmetric Standard Model,'' 
  Phys.\ Lett.\ B {\bf 134}, 429 (1984).  
  
\bibitem{Ellis:1984bm} 
  J.~R.~Ellis, C.~Kounnas and D.~V.~Nanopoulos,
  ``No Scale Supersymmetric Guts,'' 
   Nucl.\ Phys.\ B {\bf 247}, 373 (1984).  
   
\bibitem{Lahanas:1986uc} 
  A.~B.~Lahanas and D.~V.~Nanopoulos,
  ``The Road to No Scale Supergravity,'' 
   Phys.\ Rept.\  {\bf 145}, 1 (1987).  

\bibitem{Balasubramanian:2005zx} 
  V.~Balasubramanian, P.~Berglund, J.~P.~Conlon and F.~Quevedo,
  ``Systematics of moduli stabilisation in Calabi-Yau flux compactifications,''
  JHEP {\bf 0503}, 007 (2005)
  [hep-th/0502058].
  
\bibitem{Conlon:2005ki}
  J.~P.~Conlon, F.~Quevedo and K.~Suruliz,
  ``Large-volume flux compactifications: Moduli spectrum and D3/D7 soft supersymmetry breaking,''
  JHEP {\bf 0508}, 007 (2005)
  [hep-th/0505076].


 
\bibitem{Choi:2010gm} 
  K.~Choi, H.~P.~Nilles, C.~S.~Shin and M.~Trapletti,
  ``Sparticle Spectrum of Large Volume Compactification,''
  JHEP {\bf 1102}, 047 (2011)
  [arXiv:1011.0999 [hep-th]].

\bibitem{Shin:2011uk} 
  C.~S.~Shin,
  ``Anomalous U(1) Mediation in Large Volume Compactification,''
  JHEP {\bf 1201}, 084 (2012)
  [arXiv:1108.5740 [hep-ph]].
  
\bibitem{GGM} 
  C.~Cheung, A.~L.~Fitzpatrick and D.~Shih,
  ``(Extra)ordinary gauge mediation,''
  JHEP {\bf 0807}, 054 (2008)
  [arXiv:0710.3585 [hep-ph]].
  
\bibitem{Meade:2008wd}
  P.~Meade, N.~Seiberg and D.~Shih,
  ``General Gauge Mediation,''
  Prog.\ Theor.\ Phys.\ Suppl.\  {\bf 177}, 143 (2009)
  [arXiv:0801.3278 [hep-ph]].
  
\bibitem{Carpenter:2008wi}
  L.~M.~Carpenter, M.~Dine, G.~Festuccia and J.~D.~Mason,
  ``Implementing General Gauge Mediation,''
  Phys.\ Rev.\ D {\bf 79}, 035002 (2009)
  [arXiv:0805.2944 [hep-ph]].
  

  
\bibitem{Nakayama:2012zc}
  K.~Nakayama and N.~Yokozaki,
  ``Peccei-Quinn extended gauge-mediation model with vector-like matter,''  
  JHEP {\bf 1211} (2012) 158  [arXiv:1204.5420 [hep-ph]].  
  
\bibitem{Bae:2014efa} 
   K.~J.~Bae, H.~Baer, E.~J.~Chun and C.~S.~Shin,
  ``Mixed axion/gravitino dark matter from SUSY models with heavy axinos,'' 
   arXiv:1410.3857 [hep-ph].  



\bibitem{Kim:1979if} 
  J.~E.~Kim,
  ``Weak Interaction Singlet and Strong CP Invariance,''  
  Phys.\ Rev.\ Lett.\  {\bf 43}, 103 (1979).  
 
 \bibitem{Shifman:1980} 
   M.~A.~Shifman, A.~I.~Vainshtein and V.~I.~Zakharov,
  ``Can Confinement Ensure Natural CP Invariance of Strong Interactions?,''  
  Nucl.\ Phys.\ B {\bf 166}, 493 (1980).  


\bibitem{LEPchargino}
LEP2 SUSY Working Group, ALEPH, DELPHI, L3 and OPAL experiments, notes LEPSUSYWG/02-04.1 and LEPSUSYWG/01-03.1. http://lepsusy.web.cern.ch/lepsusy 

\bibitem{Dvali:1996cu} 
  G.~R.~Dvali, G.~F.~Giudice and A.~Pomarol,
  ``The Mu problem in theories with gauge mediated supersymmetry breaking,''  
  Nucl.\ Phys.\ B {\bf 478}, 31 (1996)  [hep-ph/9603238].  

\bibitem{Agashe:1999ct}
  K.~Agashe,
  ``Can multi - TeV (top and other) squarks be natural in gauge mediation?,''
  Phys.\ Rev.\ D {\bf 61} (2000) 115006
  [hep-ph/9910497].



\bibitem{Chatrchyan:2013wxa} 
  S.~Chatrchyan {\it et al.}  [CMS Collaboration],
  ``Search for gluino mediated bottom- and top-squark production in multijet final states in pp collisions at 8 TeV,''  
  Phys.\ Lett.\ B {\bf 725}, 243 (2013)  [arXiv:1305.2390 [hep-ex]].  
  
\bibitem{Aad:2013wta}
  G.~Aad {\it et al.}  [ATLAS Collaboration],
  ``Search for new phenomena in final states with large jet multiplicities and missing transverse momentum at sqrt(s)=8 TeV proton-proton collisions using the ATLAS experiment,''
    JHEP {\bf 1310}, 130 (2013)  [arXiv:1308.1841 [hep-ex]].  

\bibitem{Blumenhagen:2007sm} 
  R.~Blumenhagen, S.~Moster and E.~Plauschinn,
  ``Moduli Stabilisation versus Chirality for MSSM like Type IIB Orientifolds,''
  JHEP {\bf 0801}, 058 (2008)
  [arXiv:0711.3389 [hep-th]].
  
\bibitem{Conlon:2010ji} 
  J.~P.~Conlon and F.~G.~Pedro,
 ``Moduli Redefinitions and Moduli Stabilisation,''
  JHEP {\bf 1006}, 082 (2010)
  [arXiv:1003.0388 [hep-th]].






\end{thebibliography}
\end{document}